\documentclass[english,aps,prl,notitlepage,twocolumn,superscriptaddress,longbibliography]{revtex4-1}
\usepackage[T1]{fontenc}
\usepackage[latin9]{inputenc}
\usepackage{mathtools}
\usepackage{amsmath}
\usepackage{graphicx}

\makeatletter
\usepackage[colorlinks,citecolor=blue,linkcolor=blue]{hyperref}

\makeatother

\usepackage{babel}
\begin{document}
\title{Speeding up quantum adiabatic processes with dynamical quantum geometric
tensor}
\author{Jin-Fu Chen}
\email{chenjinfu@pku.edu.cn}

\affiliation{School of Physics, Peking University, Beijing, 100871, China}
\date{\today}
\begin{abstract}
For adiabatic controls of quantum systems, the non-adiabatic transitions
are reduced by increasing the operation time of processes. Perfect
quantum adiabaticity usually requires the infinitely slow variation
of control parameters. In this paper, we propose the dynamical quantum
geometric tensor, as a metric in the control parameter space, to speed
up quantum adiabatic processes and reach quantum adiabaticity in relatively
short time. The optimal protocol to reach quantum adiabaticity is
to vary the control parameter with a constant velocity along the geodesic
path according to the metric. For the system initiated from the $n$-th
eigenstate, the transition probability in the optimal protocol is
bounded by $P_{n}(t)\leq4\mathcal{L}_{n}^{2}/\tau^{2}$ with the operation
time $\tau$ and the quantum adiabatic length $\mathcal{L}_{n}$ induced
by the metric. Our optimization strategy is illustrated via two explicit
models, the Landau-Zener model and the one-dimensional transverse
Ising model.
\end{abstract}
\maketitle

\section{Introduction}

Optimizing the control of quantum systems is always pursued with specific
purposes in different fields, for example, to improve the fidelity
of prepared states in quantum computation \citep{Peirce1988,Caneva2009,Bason2011,Brif2014,Santos2015,Machnes2018},
and to reduce the energy dissipation in quantum thermodynamics \citep{Zulkowski2015,Solon2018,Cavina2018,Scandi2018,Vu2022}.
Adiabatic processes with time-dependent control parameters are basic
ingredients in adiabatic quantum computation \citep{Farhi2001,Sarandy2005,Nielsen2006,Menicucci2006,Aharonov2007,Albash2018}
and quantum heat engines \citep{Feldmann2000,Kieu2004,Quan_2007,Quan2009}.
A realistic adiabatic process is always completed in finite operation
time, where the non-adiabatic transition induces errors in adiabatic
quantum computation \citep{Albash2018} and consumes the output work
of a quantum heat engine \citep{Plastina2014}. The slow variation
of the Hamiltonian is thus required to reduce the non-adiabatic transition
and to reach quantum adiabaticity.

The quantum adiabatic theorem states that quantum adiabaticity is
satisfied, provided \citep{Amin2009,Sakurai2011}

\begin{equation}
\max_{t\in[0,\tau]}\left|\frac{\left\langle l(t)\right|\frac{\partial}{\partial t}\left|n(t)\right\rangle }{E_{n}(t)-E_{l}(t)}\right|\ll1,\label{eq:local1}
\end{equation}
where $\left|n(t)\right\rangle $ is the instantaneous eigenstate
of the time-dependent Hamiltonian $H(t)$ with the energy $E_{n}(t)$.
For simplicity, we assume the Hamiltonian is nondegenerate, i.e.,
for $n\ne l$, $E_{n}\ne E_{l}$. However, such a condition is insufficient
to ensure quantum adiabaticity since the overall transition probability
can still be large when plenty of eigenstates are involved during
the variation of the Hamiltonian \citep{Shevchenko2010}. Also, it
cannot directly guide the optimization of the control scheme of finite-time
adiabatic processes. To speed up quantum adiabatic processes, various
methods have been proposed, e.g., ``shortcuts to adiabaticity''
based on the inverse engineering method \citep{Demirplak2003,Demirplak2005,Masuda2008,Berry2009,Chen2010PhysRevLett104_63002,Campo2013,Santos2017,Guery-Odelin2019}
(experimental realization in \citep{Hu2018}), or the fast quasiadiabatic
method applied to few-level systems with single control parameter
\citep{Martinez-Garaot2015,Chung2017,MartinezGaraot2017,Liu2017},
yet these optimization methods require specifically designed control
schemes or are limited to specific quantum systems.

In this paper, a geometric method is proposed to optimize finite-time
adiabatic processes. Based on the high-order adiabatic approximation
method \citep{Sun_1988,Rigolin_2008,Chen2019,Chen2019a}, we formulate
a metric in the control parameter space as guidance to reduce the
non-adiabatic transition and reach quantum adiabaticity in relatively
short time. Such a metric is in a similar form to the quantum geometric
tensor \citep{Provost1980,Bengtsson2006,Zanardi2007,Venuti2007,Rezakhani2009,Rezakhani2010a},
and is thus named as ``dynamical quantum geometric tensor''. The
length $\mathcal{L}_{n}$ induced by the dynamical quantum geometric
tensor characterizes the timescale of quantum adiabaticity, and is
thus named as quantum adiabatic length. The quantum adiabatic condition
(\ref{eq:local1}) can be geometrically reformulated into

\begin{equation}
\mathcal{L}_{n}\ll\tau.\label{eq:geometric_adiabatic_condition}
\end{equation}
The optimal protocol to reach quantum adiabaticity in relatively short
time is to vary the parameter with a constant velocity along the geodesic
path according to the metric. For the $n$-th eigenstate, the transition
probability in the optimal protocol is estimated by $P_{n}(t)\approx2\mathcal{L}_{n}^{2}/\tau^{2}$
or bounded by $P_{n}(t)\leq4\mathcal{L}_{n}^{2}/\tau^{2}$ with the
operation time $\tau$. The current method is potentially helpful
to optimize finite-time adiabatic processes in experiments, e.g.,
to design control schemes for the trapped interacting Fermi gas \citep{Deng2015,Deng2018SciAdv4_5909}.

We illustrate this method via two explicit examples, the Landau-Zener
model as a two-level system \citep{1932a,Landau1932,Mullen1989,Yan2010}
and the one-dimensional transverse Ising model as a quantum many-body
system \citep{Zurek2005,Dziarmaga2005,Quan2006,Silva2008,Sachdev2017,Campo2018a,Fei2020}.
In a quantum many-body system, the quantum adiabatic length of the
path across the quantum phase transition approaches infinite in the
thermodynamic limit, which is ascribed by the divergent dynamical
quantum geometric tensor at the critical point. This relates to the
unusual finite-time scaling behavior across the quantum phase transition
\citep{Campo2018a,Fei2020,Zhang2022}, and indicates that for a many-body
system in the thermodynamic limit the quantum adiabatic condition
cannot be satisfied to cross the quantum phase transition in finite
time.

This paper is organized as follows. In Sec. \ref{sec:General-Theorem},
we propose the geometric method to optimize the control of adiabatic
processes. In Sec. \ref{sec:two-level-system}, we employ the method
for the Landau-Zener model as an illustrative example. In Sec. \ref{sec:Many-body-quantum-system},
we optimize the control for the one-dimensional transverse Ising model.
The conclusion is given in Sec. \ref{sec:Conclusion-and-discussion}.

\section{General theory\label{sec:General-Theorem}}

We propose a geometric method to optimize control schemes of finite-time
adiabatic processes for reducing the non-adiabatic transition. Due
to the external control, the system is subjected to a time-dependent
Hamiltonian $H(t)=\sum_{n}E_{n}(t)\left|n(t)\right\rangle \left\langle n(t)\right|$,
where both the energies $E_{n}(t)$ and the instantaneous eigenstate
$\left|n(t)\right\rangle $ can be time-dependent. The energies are
sorted in the increasing order $E_{0}<E_{1}<...<E_{n}<...$, and are
assumed non-degenerate, i.e., $E_{n}\ne E_{m}$ for any $n\ne m$.
The evolution of the system is governed by the time-dependent Schrödinger
equation

\begin{equation}
i\frac{\partial}{\partial t}\left|\psi(t)\right\rangle =H(t)\left|\psi(t)\right\rangle ,\label{eq:timedependengtschoridnger}
\end{equation}
We adopt a given protocol to vary the control parameter with the adjustable
operation time $\tau$.

We consider the initial state as one eigenstate of the initial Hamiltonian
$\left|\psi_{n}(0)\right\rangle =\left|n(0)\right\rangle $. The state
at time $t$ is $\left|\psi_{n}(t)\right\rangle =\sum_{l}c_{nl}(t)\left|l(t)\right\rangle $,
where the amplitudes $c_{nl}(t)$ according to Eq. (\ref{eq:timedependengtschoridnger})
satisfy 
\begin{equation}
\dot{c}_{nl}+iE_{l}c_{nl}+\sum_{m}c_{nm}\left\langle l\left|\dot{m}\right\rangle \right.=0.\label{eq:amplitude}
\end{equation}

During the evolution, the non-adiabatic transition occurs with the
probability $P_{n}(t)=\sum_{l\ne n}\left|c_{nl}(t)\right|^{2}$. Based
on the high-order adiabatic approximation method \citep{Sun_1988,Rigolin_2008},
the first-order result of the transition probability has been obtained
as \citep{Chen2019}

\begin{align}
P_{n}(t) & =\frac{1}{\tau^{2}}\sum_{l\ne n}[\left|\tilde{T}_{nl}\left(\frac{t}{\tau}\right)\right|^{2}+\left|\tilde{T}_{nl}\left(0\right)\right|^{2}-2\Lambda_{nl}(t)],\label{eq:Pn(non)}
\end{align}
where the oscillation term $\Lambda_{nl}(t)$ is 
\begin{equation}
\Lambda_{nl}(t)=-\mathrm{Re}\{e^{-i[\Phi_{n}(t)-\Phi_{l}(t)]}\tilde{T}_{nl}\left(\frac{t}{\tau}\right)\tilde{T}_{nl}^{*}\left(0\right)\},
\end{equation}
and the non-adiabatic transition rate $\tilde{T}_{nl}\left(s\right)$
is
\begin{equation}
\tilde{T}_{nl}\left(s\right)=\frac{\left\langle \tilde{l}(s)\right|\frac{\partial}{\partial s}\left|\tilde{n}(s)\right\rangle }{\tilde{E}_{n}(s)-\tilde{E}_{l}(s)},
\end{equation}
with the rescaled time $s=t/\tau$. The phase $\Phi_{n}(t)=\Phi_{n}^{\mathrm{D}}(t)+\Phi_{n}^{\mathrm{B}}(t)$
includes the dynamical phase $\Phi_{n}^{\mathrm{D}}(t)=\tau\int_{0}^{t/\tau}\tilde{E}_{n}(s)ds$
and Berry's phases $\Phi_{n}^{\mathrm{B}}(t)=-i\int_{0}^{t/\tau}\left\langle \tilde{n}(s)\right|\partial_{s}\left|\tilde{n}(s)\right\rangle ds$.
It is transparent to see that the first-order result of the probability
is bounded by $P_{n,-}(t)\leq P_{n}(t)\leq P_{n,+}(t)$ with 
\begin{equation}
P_{n,\pm}(t)=\frac{1}{\tau^{2}}\sum_{l\ne n}[\left|\tilde{T}_{nl}\left(\frac{t}{\tau}\right)\right|\pm\left|\tilde{T}_{nl}\left(0\right)\right|]^{2}.\label{eq:bounds}
\end{equation}

\begin{figure}
\includegraphics[width=8cm]{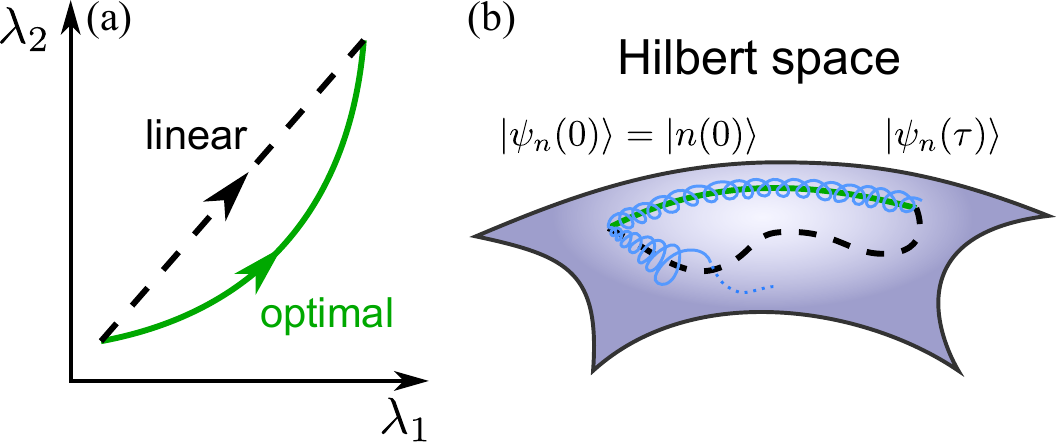}

\caption{Illustration of the evolution under the linear (black dashed line)
and the optimal (green solid curve) protocols. (a) The control protocols
with two control parameters $\lambda_{1}$ and $\lambda_{2}$. (b)
The evolution of the state $\left|\psi_{n}(t)\right\rangle $. The
instantaneous eigenstate $\left|n(t)\right\rangle $ is represented
by the black dashed and the green solid curves. In the optimal protocol,
the state $\left|\psi_{n}(t)\right\rangle $ deviates from the instantaneous
eigenstate $\left|n(t)\right\rangle $ uniformly. While in the linear
protocol, the deviation can increase greatly when the overall non-adiabatic
transition rate $\tilde{T}_{n}(s)$ becomes large. \label{fig:diagram}}
\end{figure}

We emphasize that the first-order results {[}Eqs. (\ref{eq:Pn(non)})
and (\ref{eq:bounds}){]} are only valid for slow processes when $P_{n}(t)\ll1$
is satisfied. In this situation, the state $\left|\psi_{n}(t)\right\rangle $
during the evolution is close to the instantaneous eigenstate $\left|n(t)\right\rangle $.
With the shorter operation time, the first-order approximation may
fail at a specific time point when the overall non-adiabatic transition
rate $\tilde{T}_{n}(s)\coloneqq[\sum_{l\ne n}|\tilde{T}_{nl}(s)|^{2}]^{1/2}$
becomes large. To make the quantum adiabatic condition (\ref{eq:local1})
possibly hold on the whole evolution, the optimal protocol to vary
the control parameter is to keep 
\begin{equation}
\tilde{T}_{n}(s)=\mathrm{const}.\label{eq:constantoverallnon-adiabaticrate}
\end{equation}
We illustrate the evolution of the state under the linear and the
optimal protocols in Fig. \ref{fig:diagram}. In the linear protocol,
the state $\left|\psi_{n}(t)\right\rangle $ deviates from the instantaneous
eigenstate $\left|n(t)\right\rangle $ increasingly, and the final
state $\left|\psi_{n}(\tau)\right\rangle $ becomes much different
from the final instantaneous eigenstate $\left|n(\tau)\right\rangle $.
In the optimal protocol, the transition probability $P_{n}(t)$ is
regularly oscillated for a few-level system (Sec. \ref{sec:two-level-system}),
and becomes uniform for a quantum many-body system (Sec. \ref{sec:Many-body-quantum-system}).
One can thus properly control the deviation from the instantaneous
eigenstate $\left|n(t)\right\rangle $.

To estimate the transition probability, we define the quantum adiabatic
length $\mathcal{L}_{n}$ for the $n$-th eigenstate with the overall
non-adiabatic transition rate $\tilde{T}_{n}(s)$ as

\begin{equation}
\mathcal{L}_{n}\coloneqq\int_{0}^{1}\tilde{T}_{n}(s)ds.\label{eq:quantumadiabaticlength}
\end{equation}
We consider the variation of the Hamiltonian $H(t)=H[\vec{\lambda}(t)]$
through multiple control parameters $\vec{\lambda}=\{\lambda_{i}\}$.
The quantum adiabatic length is determined by the path in the control
parameter space
\begin{equation}
\mathcal{L}_{n}=\int_{0}^{1}\sqrt{\sum_{ij}\tilde{\lambda}_{i}^{\prime}(s)g_{n,ij}(\vec{\lambda})\tilde{\lambda}_{j}^{\prime}(s)}ds,
\end{equation}
and is independent of the control protocol on the path. We coin the
dynamical quantum geometric tensor for the metric

\begin{equation}
g_{n,ij}(\vec{\lambda})=\mathrm{Re}\sum_{l\ne n}\frac{\left\langle l\right|\frac{\partial H}{\partial\lambda_{i}}\left|n\right\rangle \left\langle n\right|\frac{\partial H}{\partial\lambda_{j}}\left|l\right\rangle }{(E_{n}-E_{l})^{4}},\label{eq:quantumadiabaticmetric}
\end{equation}
due to its similarity to the quantum geometric tensor \citep{Provost1980}
except that the index in the numerator is $4$ instead of $2$. In
the optimal protocol, the transition probability according to Eq.
(\ref{eq:Pn(non)}) is estimated by
\begin{equation}
P_{n}(t)\approx\frac{2\mathcal{L}_{n}^{2}}{\tau^{2}},\label{eq:estimatefromadiabaticlength}
\end{equation}
when neglecting the oscillation term. One can further choose the geodesic
path connecting $\vec{\lambda}(0)$ and $\vec{\lambda}(\tau)$ to
minimize the quantum adiabatic length $\mathcal{L}_{n}$ and reduce
the transition probability $P_{n}(t)$. Take into account the oscillation
term $\Lambda_{nl}(t)$, the upper bound (\ref{eq:bounds}) of the
transition probability for the optimal protocol becomes $P_{n,+}(t)=4\mathcal{L}_{n}^{2}/\tau^{2}.$
The quantum adiabatic length $\mathcal{L}_{n}$, with the dimension
of time, indicates the timescale of quantum adiabaticity, and the
quantum adiabatic condition is geometrically reformulated in Eq. (\ref{eq:geometric_adiabatic_condition}).

The proposed dynamical quantum geometric tensor fairly assesses the
non-adiabatic transition from $\left|n\right\rangle $ to all the
other states. In Ref. \citep{Rezakhani2009}, the used metric for
the optimization is an approximation of Eq. (\ref{eq:quantumadiabaticmetric})
by substituting all $E_{n}-E_{l}$ in Eq. (\ref{eq:quantumadiabaticmetric})
with the energy gap between the ground state and the first excited
state. With the dynamical quantum geometric tensor, the optimization
of the protocol to reach quantum adiabaticity in relatively short
time is converted to finding the geodesic path on the control parameter
space.

\section{Landau-Zener Model\label{sec:two-level-system}}

We employ the above geometric method to optimize the control for the
simplest quantum system, i.e., a two-level system, which also serves
as the basic element as a qubit in quantum computation. The precise
control of the state of the qubit ensures the reliability of a quantum
computer \citep{Albash2018}. We consider the well-known Landau-Zener
model \citep{1932a,Landau1932} described by the Hamiltonian

\begin{equation}
H=\frac{\Delta}{2}(\sigma_{x}+\lambda\sigma_{z}),\label{eq:Hamiltonian_twolevel}
\end{equation}
where $\lambda$ serves as the control parameter, and $\sigma_{x},\sigma_{z}$
are the Pauli matrices. The origin Landau-Zener model adopts a linear
protocol to vary the control parameter $\lambda$.\textbf{ }The initial
state is chosen as the ground state $\left|\psi_{\mathrm{g}}(0)\right\rangle =\left|\mathrm{g}(0)\right\rangle $
with the initial control parameter satisfying $\left|\lambda\right|\gg1$.
For long operation time, the transition probability $P_{\mathrm{g}}=\left|\left\langle \mathrm{e}(\tau)\left|\psi_{\mathrm{g}}(\tau)\right\rangle \right.\right|^{2}$
approaches zero at the end of the evolution.

To derive the optimal protocol, we rewrite the Hamiltonian {[}Eq.
(\ref{eq:Hamiltonian_twolevel}){]} into

\begin{equation}
H=\frac{\Delta}{2}\sqrt{1+\lambda^{2}}\left(\left|\mathrm{e}\right\rangle \left\langle \mathrm{e}\right|-\left|\mathrm{g}\right\rangle \left\langle \mathrm{g}\right|\right),
\end{equation}
with the instantaneous eigenstates

\begin{align}
\left|\mathrm{g}\right\rangle  & =\left(\begin{array}{c}
-\sqrt{\frac{\sqrt{1+\lambda^{2}}-\lambda}{2\sqrt{1+\lambda^{2}}}}\\
\sqrt{\frac{\sqrt{1+\lambda^{2}}+\lambda}{2\sqrt{1+\lambda^{2}}}}
\end{array}\right),\;\left|\mathrm{e}\right\rangle =\left(\begin{array}{c}
\sqrt{\frac{\sqrt{1+\lambda^{2}}+\lambda}{2\sqrt{1+\lambda^{2}}}}\\
\sqrt{\frac{\sqrt{1+\lambda^{2}}-\lambda}{2\sqrt{1+\lambda^{2}}}}
\end{array}\right).
\end{align}
According to Eq. (\ref{eq:constantoverallnon-adiabaticrate}), the
optimal protocol satisfies

\begin{equation}
\frac{[\tilde{\lambda}_{\mathrm{op}}^{\prime}(s)]^{2}}{[1+\tilde{\lambda}_{\mathrm{op}}(s)^{2}]^{3}}=\mathrm{const}.
\end{equation}
With the initial and the final values of the control parameter $\tilde{\lambda}(0)=-\lambda_{0}$
and $\tilde{\lambda}(1)=\lambda_{0}$, the optimal protocol is solved
as

\begin{equation}
\tilde{\text{\ensuremath{\lambda}}}_{\mathrm{op}}(s)=\frac{-\lambda_{0}(1-2s)}{\sqrt{1+4\lambda_{0}^{2}s(1-s)}},\label{eq:LZspecialprotocol}
\end{equation}
while the linear protocol is $\tilde{\text{\ensuremath{\lambda}}}_{\mathrm{lin}}(s)=-\lambda_{0}(1-2s)$.
For the two-level system, the quantum adiabatic lengths are identical
for the ground and the excited states, i.e., $\mathcal{L}_{\mathrm{g}}=\mathcal{L}_{\mathrm{e}}=\left|\lambda_{0}\right|/(\Delta\sqrt{\lambda_{0}^{2}+1})$.

\begin{figure*}[t]
\includegraphics[width=12cm]{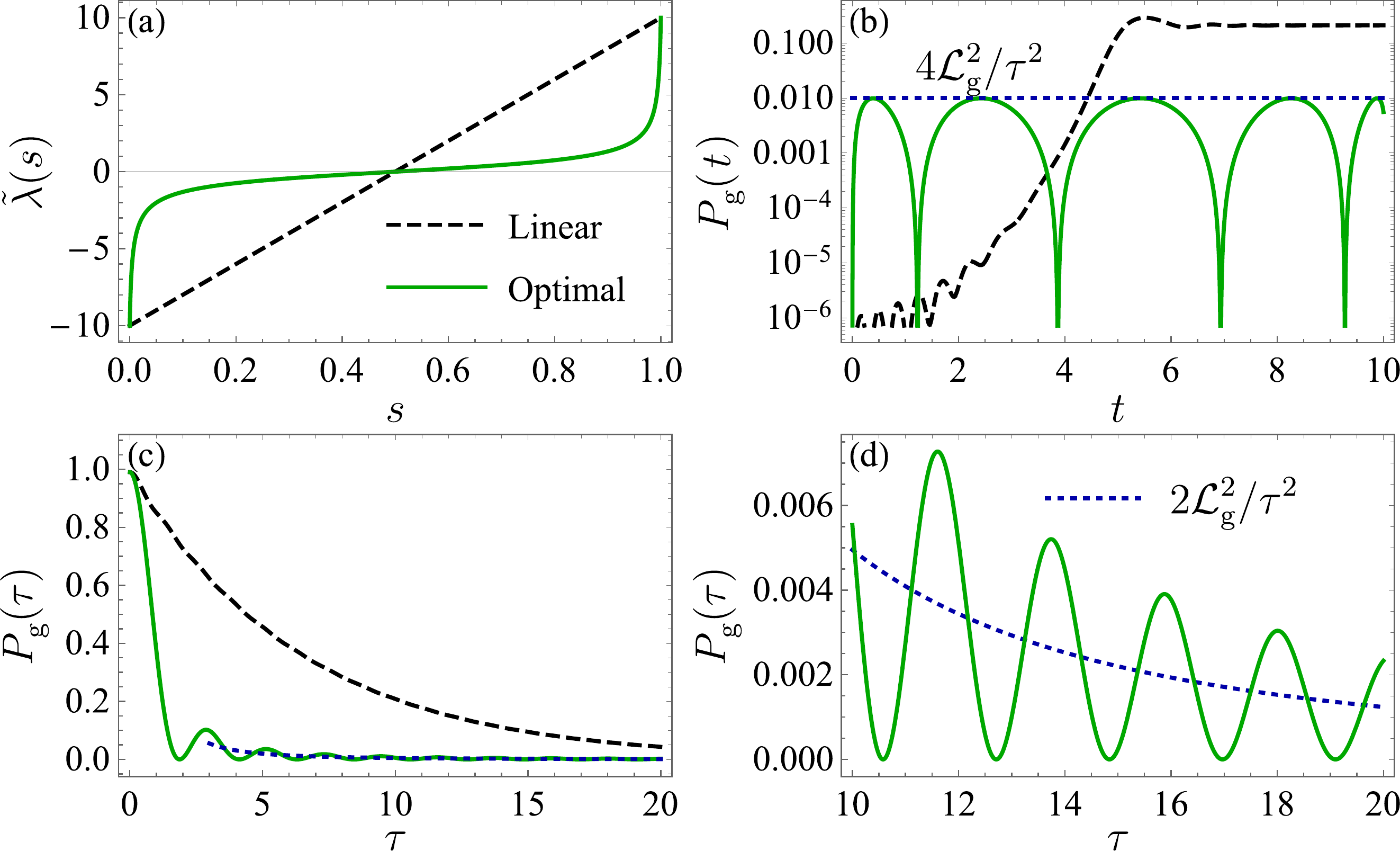}

\caption{The transition probability for the Landau-Zener model. The parameters
are set to be $\lambda_{0}=10$ and $\Delta=2$. (a) The control schemes
for the linear and the optimal protocols. (b) The transition probability
$P_{\mathrm{g}}(t)$ during the whole evolution with $\tau=10$. The
blue dotted line shows the upper bound $P_{\mathrm{g}}(t)\protect\leq4\mathcal{L}_{\mathrm{g}}^{2}/\tau^{2}$
for the optimal protocol. (c) and (d) The final transition probability
$P_{\mathrm{g}}(\tau)$ as a function of the operation time $\tau$.
In the optimal protocol, the final transition probability is estimated
by $P_{\mathrm{g}}(\tau)\approx2\mathcal{L}_{\mathrm{g}}^{2}/\tau^{2}$
(blue dotted curve) with $\mathcal{L}_{\mathrm{g}}=0.498$. \label{fig:laudauzener}}
\end{figure*}

Figure \ref{fig:laudauzener} shows the numerical results of the transition
probability $P_{\mathrm{g}}(\tau)$ for the Landau-Zener model under
the linear and the optimal protocols. In Fig. \ref{fig:laudauzener}(a),
we compare the linear protocol $\tilde{\text{\ensuremath{\lambda}}}_{\mathrm{lin}}(s)$
and the optimal protocol $\tilde{\lambda}_{\mathrm{op}}(s)$ with
$\lambda_{0}=10$. In the optimal protocol, the control parameter
$\lambda$ is varied fast (slowly) with large (small) energy spacing
at $s=0$ and $1$ ($s=0.5$). Figure. \ref{fig:laudauzener}(b) shows
the transition probability $P_{\mathrm{g}}(t)$ of the two protocols
during the whole evolution with $\tau=10$ and $\Delta=2$. In the
linear protocol (black dashed curve), the transition probability $P_{\mathrm{g}}(t)$
keeps increasing before the energy spacing reaches the minimum at
$t=\tau/2$. In the optimal protocol (green solid curve), the transition
probability $P_{\mathrm{g}}(t)$ increases rapidly at the initial
time, but soon saturates the upper bound $4\mathcal{L}_{\mathrm{g}}^{2}/\tau^{2}$
(blue dotted line). We observe the oscillation in the transition probability
$P_{\mathrm{g}}(t)$. Its value approaches almost zero at specific
moments. Such a phenomenon can be understood from the first-order
result Eq. (\ref{eq:Pn(non)}). For the two-level system, there is
only one term $l=\mathrm{e}$ left in the summation in Eq. (\ref{eq:Pn(non)}),
and the transition probability $P_{\mathrm{g}}(t)$ can approach zero
with a proper value of the phase factor in the oscillation term $\Lambda_{\mathrm{ge}}(t)$.
The oscillation phenomenon has also been observed in the quantum harmonic
oscillator with the time-dependent frequency \citep{Chen2019a}. 

In Fig \ref{fig:laudauzener}(c) and (d), we compare the final transition
probability $P_{\mathrm{g}}(\tau)$ of the two protocols with different
operation time $\tau$. In the optimal protocol, the probability $P_{\mathrm{g}}(\tau)$
decreases more rapidly (green curve) with the increase of the operation
time, and is estimated by $P_{\mathrm{g}}(t)\approx2\mathcal{L}_{\mathrm{g}}^{2}/\tau^{2}$
with neglecting the oscillation. The quantum adiabaticity is reached
with shorter operation time in the optimal protocol than in the linear
protocol.

In Appendix \ref{sec:General-two-level-system}, we optimize the control
of a general two-level system with changing the direction of the control
parameters.

\section{One-dimensional transverse Ising model\label{sec:Many-body-quantum-system}}

It is intriguing to employ the geometric method to optimize the control
of quantum many-body systems. For a system with multiple energy eigenstates,
the non-adiabatic transitions to all the other states contribute to
the transition probability $P_{n}(t)$, whose behavior can still be
investigated from the dynamical quantum geometric tenser. As an illustrative
example, we consider the one-dimensional transverse Ising model \citep{Sachdev2017}.
The Hamiltonian reads

\begin{equation}
H=-J\sum_{i=1}^{N}\left(\sigma_{i}^{z}\sigma_{i+1}^{z}+\lambda\sigma_{i}^{x}\right).
\end{equation}
We consider the site number $N$ even and periodic boundary condition
$\sigma_{N+1}=\sigma_{1}$. The sign of $J$ does not affect the results
of the transition probability $P_{n}(t)$, and we set $J=1$ in all
the numerical calculation for convenience. This model can be mapped
into a free Fermion model described by quasiparticles, and is thus
fully solvable. The quantum phase transition of this model occurs
at the critical points $\lambda=\pm1$ \citep{Sachdev2017}. The external
field $\lambda$ serves as the control parameter, the control scheme
of which is usually considered as the instant \citep{Quan2006,Silva2008}
or the linear quenches \citep{Zurek2005,Dziarmaga2005}. For the linear
quench across the critical point, the average excitation \citep{Campo2018a}
and the average excess work \citep{Fei2020} scale with the operation
time as $\tau^{-1/2}$.

For the one-dimensional transverse Ising model in the thermodynamic
limit, the quantum phase transition close the energy gap of the system
at the critical points, resulting in the divergence of the quantum
geometry tensor \citep{Venuti2007,Zanardi2007}. The divergence also
exists for the dynamical quantum geometric tensor, and prevents constructing
an optimal protocol to cross the critical point, but the current method
can be used to optimize the control scheme either for a finite-size
system or without crossing the critical point.

Under the Jordan-Wigner transformation, the model is mapped to a free
Fermion model with the Hamiltonian \citep{Sachdev2017}

\begin{align}
H & =\sum_{k>0}H_{k},\label{eq:27}
\end{align}
where $k$ ranges from $0$ to $\pi-2\pi/N$ with the interval $2\pi/N$.
In the $k$-subspace, the Hamiltonian entangles the modes $k$ and
$-k$ as

\begin{equation}
H_{k}=2J\psi_{k}^{\dagger}\left(\begin{array}{cc}
\lambda-\cos k & -i\sin k\\
i\sin k & -\lambda+\cos k
\end{array}\right)\psi_{k},\label{eq:Hkpair}
\end{equation}
in terms of $\psi_{k}^{\dagger}=\left(\begin{array}{cc}
c_{k}^{\dagger} & c_{-k}\end{array}\right)$. For the mode $k=0$ or $\pi$, the evolution can be also described
by Eq. (\ref{eq:Hkpair}) with $\psi_{0}^{\dagger}=\left(\begin{array}{cc}
c_{0}^{\dagger} & c_{\pi}\end{array}\right)$, and the two modes do not mix since the off-diagonal terms are zero.
The Hamiltonian is diagonalized under the Bogliubov transformation
as

\begin{equation}
H_{k}=\epsilon_{k}(A_{k}^{\dagger}A_{k}-\frac{1}{2}).
\end{equation}
The energy and the annihilation operator of the quasiparticle are
$\epsilon_{k}=2J\left(\lambda^{2}-2\lambda\cos k+1\right)^{1/2}$
and $A_{k}=u_{k}c_{k}-iv_{k}c_{-k}^{\dagger}$, where the coefficients
are $u_{k}=\cos(\theta_{k}/2)$ and $v_{k}=\sin(\theta_{k}/2)$ with
$\tan\theta_{k}=\sin k/(\lambda-\cos k)$.

For the initial ground state, the wave-function between different
pairs $\pm k$ are in the direct product form. We write down the ground-state
wave-function in each $k$-subspace as

\begin{equation}
\left|\mathrm{g}(k)\right\rangle =u_{k}\left|0_{k}0_{-k}\right\rangle +iv_{k}\left|1_{k}1_{-k}\right\rangle ,
\end{equation}
where $\left|l_{k}\right\rangle $ is the Fock state satisfying $c_{k}^{\dagger}c_{k}\left|l_{k}\right\rangle =l_{k}\left|l_{k}\right\rangle $
with $l=\pm1$. The single-occupy states are always the eigenstates
$H_{k}\left|0_{k}1_{-k}\right\rangle =0,$ $H_{k}\left|1_{k}0_{-k}\right\rangle =0$
of the Hamiltonian $H_{k}$. The finite-time variation does not induce
the non-adiabatic transition to these states. Therefore, the Hamiltonian
in each $k$-subspace is equivalent to that of a two-level system.
The non-adiabatic transitions are obtained with several pairs of states
$\left|0_{k}0_{-k}\right\rangle $ and $\left|1_{k}1_{-k}\right\rangle $.

We employ the geometric method to optimize the control scheme of the
quench for the one-dimensional transverse Ising model with finite
site number $N$. Our task is to find the optimal protocol to vary
the external field $\lambda$. As shown in Appendix \ref{sec:transverseisingmodel},
the quantum adiabatic length is obtained as

\begin{equation}
d\mathcal{L}_{\mathrm{g}}=\sum_{k>0}\frac{\sin k}{8J\left(\lambda^{2}-2\lambda\cos k+1\right)^{3/2}}d\lambda,\label{eq:length}
\end{equation}
where the summation of $k$ is limited to $k=2\pi/N,4\pi/N,...\pi-2\pi/N$.
The optimal protocol $\tilde{\lambda}_{\mathrm{op}}(s)$ follows as

\begin{equation}
[\tilde{\lambda}_{\mathrm{op}}^{\prime}(s)]^{2}\sum_{k>0}\frac{\sin^{2}k}{\left[\tilde{\lambda}_{\mathrm{op}}(s)^{2}-2\tilde{\lambda}_{\mathrm{op}}(s)\cos k+1\right]^{3}}=\mathrm{const}.\label{eq:optimalisingmodel}
\end{equation}
Due to the quasiparticle representation of this model, the transition
probability $P_{\mathrm{g}}(t)$ of the ground state of this many-body
system is the product of the transition probabilities of the two-level
system in each $k$-subspace. 

\begin{figure}
\includegraphics[width=6cm]{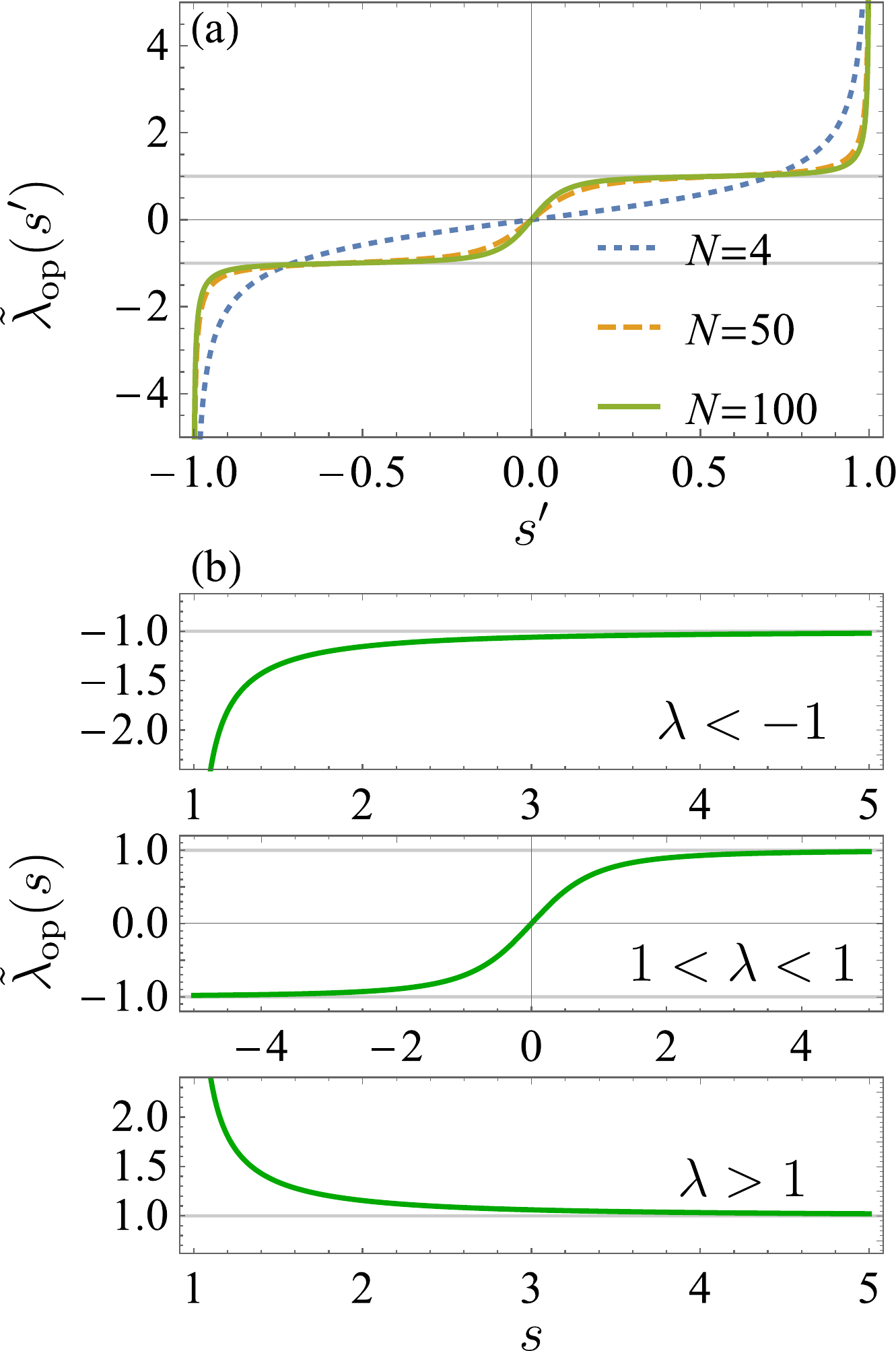}

\caption{The optimal protocol for the one-dimensional transverse Ising model.
(a) The optimal protocols for finite site numbers $N=4,50,100$ with
the time $s^{\prime}$ rescaled to $(-1,1)$. (b) The optimal protocol
in the thermodynamic limit $N\rightarrow\infty$. We show the the
optimal protocols Eq. (\ref{eq:lambds}) for the three regions $\lambda<-1$,
$-1<\lambda<1$ and $\lambda>1$. The metric is divergent at the critical
points $\lambda=\pm1$ (horizontal gray line), which cannot be crossed
in finite time. \label{fig:The-special-protocolfiniteN}}
\end{figure}

For given site number $N$, the optimal protocol can be numerically
solved by Eq. (\ref{eq:optimalisingmodel}). For $N=4$, only one
term with $k=\pi/2$ leaves in the summation, and the optimal protocol
$\tilde{\lambda}_{\mathrm{op}}(s^{\prime})=s^{\prime}/\sqrt{1-s^{\prime2}}$
coincides with that of the Landau-Zener model with a rescaled time
$s^{\prime}\in(-1,1)$. In Fig. \ref{fig:The-special-protocolfiniteN}
(a), the optimal protocols are shown for different site numbers $N=4,50,100$.
With the increase of the site number $N$, it consumes more operation
time to cross the critical points $\lambda=\pm1$.

In the thermodynamic limit $N\rightarrow\infty$, Eq. (\ref{eq:optimalisingmodel})
is simplified into

\begin{equation}
\frac{\tilde{\lambda}_{\mathrm{op}}^{\prime}(s)^{2}}{\left|\tilde{\lambda}_{\mathrm{op}}(s)^{2}-1\right|^{3}}=\mathrm{const},
\end{equation}
and the optimal protocol is explicitly obtained as

\begin{equation}
\tilde{\lambda}_{\mathrm{op}}(s)=\begin{cases}
\frac{-s}{\sqrt{s^{2}-1}} & \lambda<-1\\
\frac{s}{\sqrt{1+s^{2}}} & -1<\lambda<1\\
\frac{s}{\sqrt{s^{2}-1}} & \lambda>1,
\end{cases}\label{eq:lambds}
\end{equation}
as shown in Fig. \ref{fig:The-special-protocolfiniteN}(b). The constant
has been absorbed into the rescaled time $s$ here. In the three regions
$\lambda<-1$, $-1<\lambda<1$ and $\lambda>1$ of the control parameter,
the ranges of the rescaled time are $s\in(1,+\infty)$, $(-\infty,+\infty)$
and $(1,+\infty)$, respectively. Equation (\ref{eq:lambds}) shows
that in the thermodynamic limit $N\rightarrow\infty$ the optimal
protocol cannot cross the critical points $\lambda=\pm1$ in any finite
time process.

\begin{figure*}[t]
\includegraphics[width=12cm]{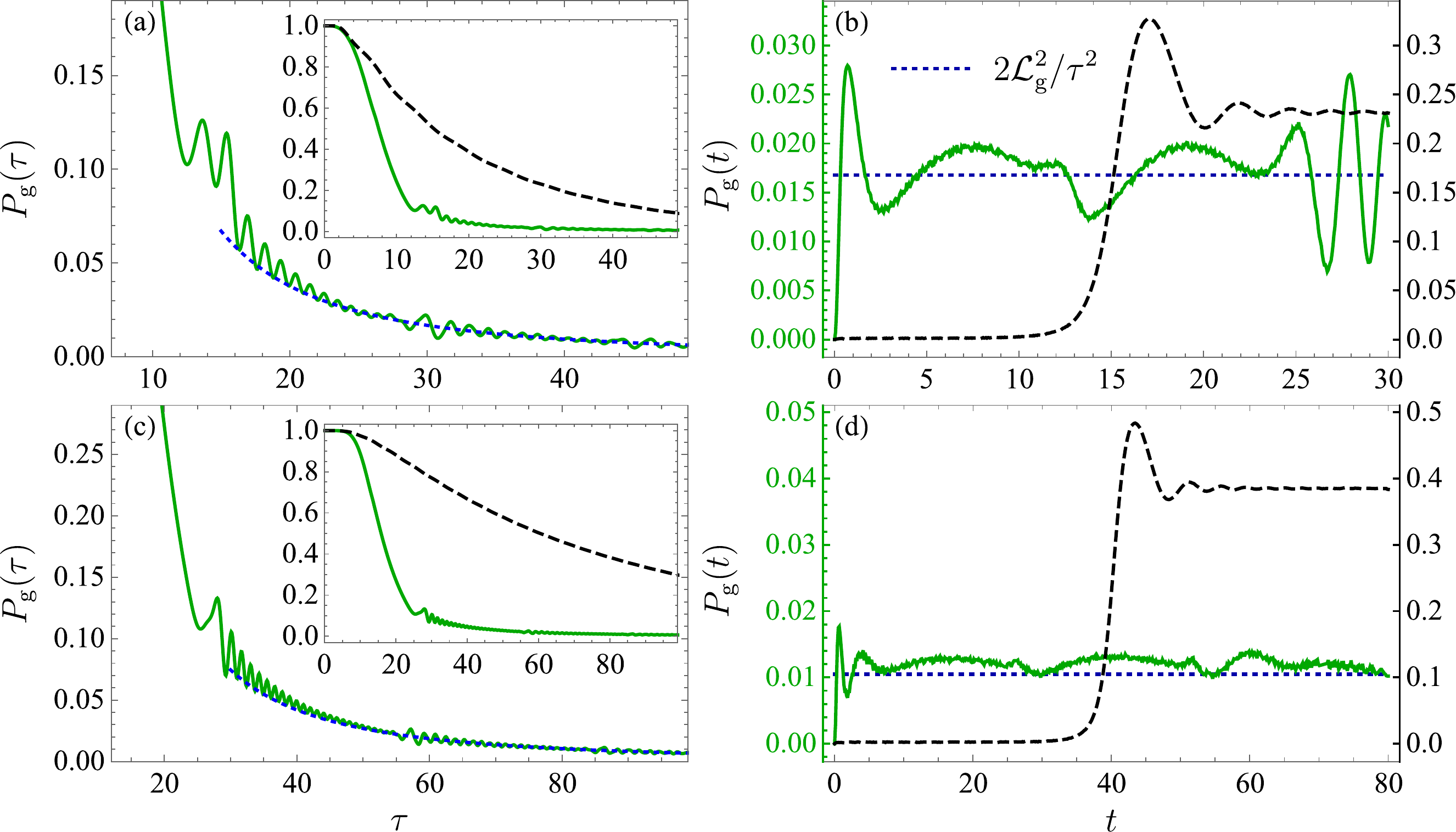}\caption{The transition probability for the one-dimensional transverse Ising
model. The site number is $N=50$ in (a), (b) and $N=100$ in (c),
(d) with $J=1$. The external field is varied from $\tilde{\lambda}(0)=2$
to $\tilde{\lambda}(1)=0$ via the optimal protocol $\tilde{\lambda}_{\mathrm{op}}(s)$
(green solid curve) and the linear protocol $\tilde{\lambda}_{\mathrm{lin}}(s)=2(1-s)$
(black dashed curve). All the blue dashed curves show the estimation
$2\mathcal{L}_{\mathrm{g}}^{2}/\tau^{2}$. (a) and (c) The final transition
probability $P_{\mathrm{g}}(\tau)$ as a function of the operation
time $\tau$. The results of the linear protocol are shown in the
insets. (b) and (d) The transition probability $P_{\mathrm{g}}(t)$
during the whole evolution. The values of $P_{\mathrm{g}}(t)$ in
the linear protocol are multiplied ten times. \label{fig:numericalresultoftransverseising}}
\end{figure*}

Figure \ref{fig:numericalresultoftransverseising} shows the numerical
results of the transition probability $P_{\mathrm{g}}(\tau)$ with
the linear protocol $\tilde{\lambda}_{\mathrm{lin}}(s)=\tilde{\lambda}(0)(1-s)+\tilde{\lambda}(1)s$
and the optimal protocol $\tilde{\lambda}_{\mathrm{op}}(s)$ {[}Eq.
(\ref{eq:optimalisingmodel}){]} for the one-dimensional transverse
Ising model. The initial and the final values of the control parameter
are $\tilde{\lambda}(0)=2$ and $\tilde{\lambda}(1)=0$. The site
number is $N=50$ in (a), (b) and $N=100$ in (c), (d). The transition
probability of the ground state is obtained by numerically solving
the time-dependent Schrödinger equation with $J=1$. Figure \ref{fig:numericalresultoftransverseising}(a)
and (c) present the final transition probability $P_{\mathrm{g}}(\tau)$
as a function of the operation time $\tau$. The final transition
probability in the optimal protocol is well estimated by $P_{\mathrm{g}}(\tau)\approx2\mathcal{L}_{\mathrm{g}}^{2}/\tau^{2}$
(blue dotted curve), and is much smaller than that in the linear protocol
as shown by the insets. With more spins in the system, it requires
a longer operation time to remain the same transition probability
to cross the critical point $\lambda=1$. The phenomenon is induced
by the modes around $k\simeq0$ with slower dynamics \citep{Campo2018a}.

Figure \ref{fig:numericalresultoftransverseising}(b) and (d) present
the transition probability $P_{\mathrm{g}}(t)$ during the whole evolution
with given operation time $\tau=30$ and $\tau=80$, respectively.
In the linear protocol, the transition probability $P_{\mathrm{g}}(t)$
increase rapidly at the moment $t/\tau=0.5$ across the critical point.
In the optimal protocol, the transition probability during the whole
evolution is well estimated by $P_{\mathrm{g}}(t)\approx2\mathcal{L}_{\mathrm{g}}^{2}/\tau^{2}$
(blue dotted line). The oscillation in $P_{\mathrm{g}}(t)$ is much
weaker but more irregular compared to the case of the two-level system,
since $P_{\mathrm{g}}(t)$ is the product of the transition probabilities
in each $k$-subspace.

\section{Conclusion\label{sec:Conclusion-and-discussion}}

We proposed the dynamical quantum geometric tensor to speed up finite-time
adiabatic processes. The dynamical quantum geometric tensor is a metric
in the control parameter space. The length induced by metric, i.e.,
the quantum adiabatic length, determines the timescale of quantum
adiabaticity. The optimal protocol is to vary the control parameter
with a constant velocity along the geodesic path according to the
metric, and the transition probability is estimated (bounded) by the
quantum adiabatic length as $P_{n}\approx2\mathcal{L}_{n}^{2}/\tau^{2}$
($P_{n}\leq4\mathcal{L}_{n}^{2}/\tau^{2}$). We employ the geometric
method to optimize the control of the Landau-Zener model and the one-dimensional
transverse Ising model, and verify the transition probability in the
optimal protocol is much smaller than that in the linear protocol
with given operation time.
\begin{acknowledgments}
J.F. Chen thanks C.P. Sun, Hui Dong, and Zhaoyu Fei in Graduate School
of China Academy of Engineering Physics for helpful discussions. This
work is supported by the National Natural Science Foundation of China
(NSFC) under Grants No. 11775001, No. 11825501, and No. 12147157.
\end{acknowledgments}

\appendix

\section{General two-level system\label{sec:General-two-level-system}}

For a two-level system, the Hamiltonian is generally written as

\begin{equation}
H=\frac{1}{2}(\lambda_{x}\sigma_{x}+\lambda_{y}\sigma_{y}+\lambda_{z}\sigma_{z}),
\end{equation}
with the control parameters $\vec{\lambda}=(\lambda_{x},\lambda_{y},\lambda_{z})$.
According to Eq. (\ref{eq:quantumadiabaticmetric}), the dynamical
quantum geometric tensor is obtained as

\begin{equation}
g_{\mathrm{g}}(\vec{\lambda})=\frac{1}{4\lambda^{6}}\left(\begin{array}{ccc}
\lambda_{y}^{2}+\lambda_{z}^{2} & -\lambda_{x}\lambda_{y} & -\lambda_{x}\lambda_{z}\\
-\lambda_{x}\lambda_{y} & \lambda_{x}^{2}+\lambda_{z}^{2} & -\lambda_{y}\lambda_{z}\\
-\lambda_{x}\lambda_{z} & -\lambda_{y}\lambda_{z} & \lambda_{x}^{2}+\lambda_{y}^{2}
\end{array}\right),
\end{equation}
with $\lambda=\sqrt{\lambda_{x}^{2}+\lambda_{y}^{2}+\lambda_{z}^{2}}$.
For the two-level system, the metric for the excited state is the
same $g_{\mathrm{e}}(\vec{\lambda})=g_{\mathrm{g}}(\vec{\lambda})$.
Under the sphere coordinates $(\lambda,\theta,\phi)$ with $\cos\theta=\lambda_{z}/\lambda$
and $\tan\phi=\lambda_{y}/\lambda_{x}$, the quantum adiabatic length
is simplified into

\begin{equation}
d\mathcal{L}_{\mathrm{g}}^{2}=\frac{\sin^{2}\theta d\theta^{2}+d\phi^{2}}{4\lambda^{2}}.\label{eq:lengthg}
\end{equation}
The metric is degenerate along the direction $\vec{\lambda}/\lambda$,
since the changing strength with fixed direction does not generate
the transition between different eigenstates. We constrain the control
of the parameters on a sphere $\lambda=\mathrm{const}$. The geodesic
paths on the sphere are large circles.

\begin{figure}
\includegraphics[width=6cm]{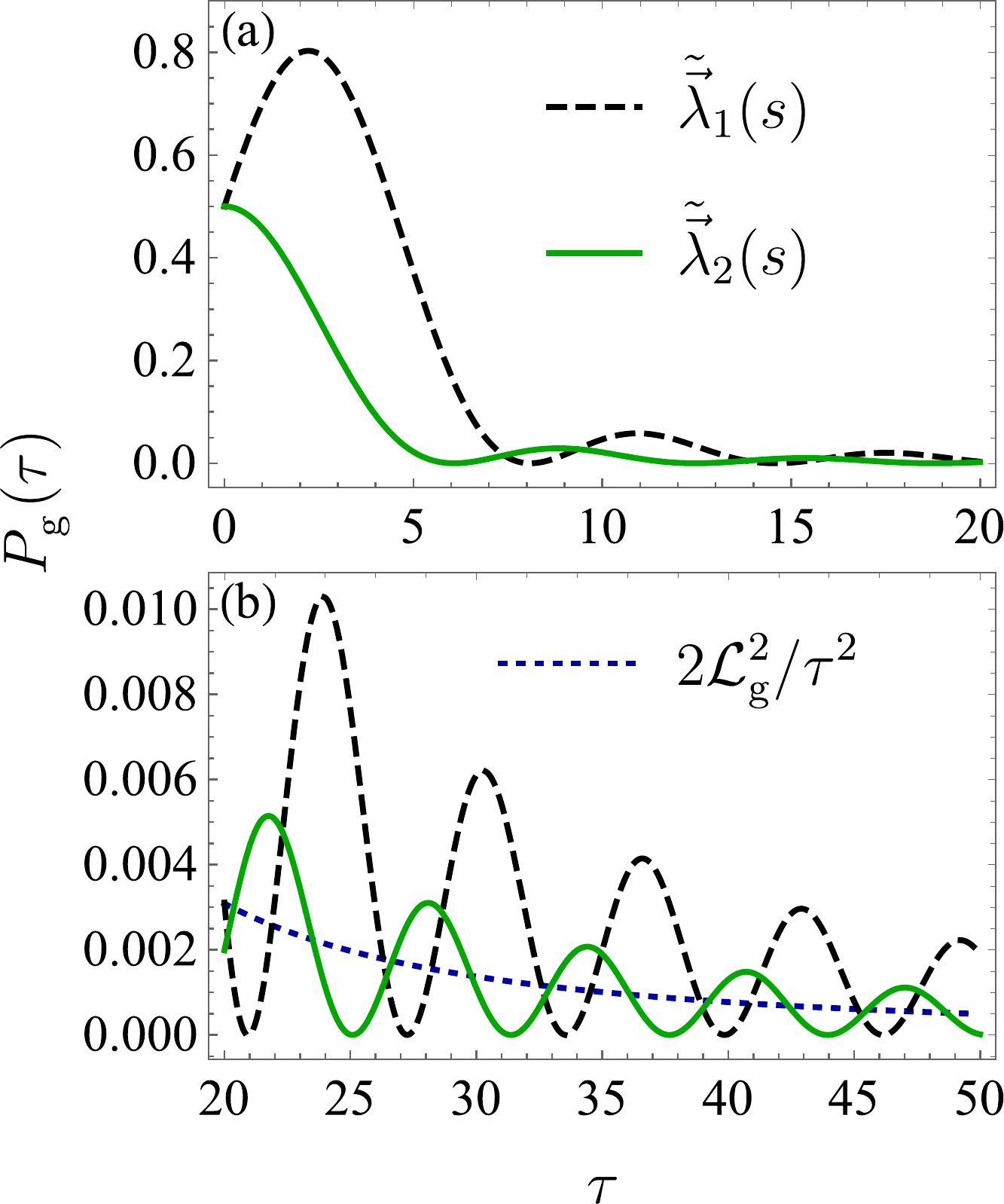}

\caption{The final transition probability $P_{\mathrm{g}}(\tau)$ for the control
of the external field $\vec{\lambda}$ constraint on the sphere $\lambda=1$.
Two protocols are considered, the small-circle protocol $\tilde{\vec{\lambda}}_{1}(s)$
and the large-circle protocol $\tilde{\vec{\lambda}}_{2}(s)$. (a)
The results with short operation time $\tau\in(0,20]$. (b) The results
with longer operation time $\tau\in[20,50]$. The blue dotted curve
shows the estimation $P_{\mathrm{g}}(\tau)\approx2\mathcal{L}_{\mathrm{g}}^{2}/\tau^{2}$
with the length $\mathcal{L}_{\mathrm{g}}=\pi/4$ of the path on the
large circle.\label{fig:spherecontrol}}
\end{figure}

We next compare different control protocols to vary the external field
constraint on the sphere $\lambda=1$. Two protocols are adopted to
vary the external field from the initial point $\tilde{\vec{\lambda}}(0)=(1/\sqrt{2},0,1/\sqrt{2})$
to the final point $\tilde{\vec{\lambda}}(1)=(-1/\sqrt{2},0,1/\sqrt{2})$,
with one on a small circle

\begin{equation}
\tilde{\vec{\lambda}}_{1}(s)=\frac{\sqrt{2}}{2}(\cos(\pi s),\sin(\pi s),1),
\end{equation}
and the other on a large circle (geodesic path)

\begin{equation}
\tilde{\vec{\lambda}}_{2}(s)=(\sin[\frac{\pi}{4}(1-2s)],0,\cos[\frac{\pi}{4}(1-2s)]).
\end{equation}
According to Eq. (\ref{eq:lengthg}), the quantum adiabatic lengths
of the two paths are $\mathcal{L}_{\mathrm{g}}=\sqrt{2}\pi/4$ and
$\pi/4$. In Figure \ref{fig:spherecontrol}, we show the transition
probability $P_{\mathrm{g}}(\tau)$ for the two protocols with different
operation time $\tau$. The transition probability $P_{\mathrm{g}}(\tau)$
of the protocol on the geodesic path is smaller. In both protocols,
$P_{\mathrm{g}}(\tau)$ can be estimated by $P_{\mathrm{g}}(\tau)\approx2\mathcal{L}_{\mathrm{g}}^{2}/\tau^{2}$
when neglecting the oscillation. In Fig. \ref{fig:spherecontrol}(b),
the estimation of the quantum adiabatic length is shown for the second
protocol by the blue dotted curve.

\begin{widetext}

\section{Optimal protocol of the one-dimensional transverse Ising model\label{sec:transverseisingmodel}}

For the one-dimensional transverse Ising model, we represent the instantaneous
eigenstates of the many-body system as the tensor product $\left|\Psi_{\{n(k)\}}\right\rangle =\underset{k>0}{\bigotimes}\left|n(k)\right\rangle $,
where $\left|n(k)\right\rangle $ is the eigenstate in each subspace,
and $n=\mathrm{g}$ and $\mathrm{e}$ represents the ground state
and the excited state, respectively. Here, we do not consider the
modes $k\ne0,\pi$, since the eigenstates of them remain unchanged
when varying the control parameter. The quantum adiabatic length $\mathcal{L}_{\{n(k)\}}$
of the eigenstate $\left|\Psi_{\{n(k)\}}\right\rangle $ is determined
by Eq. (\ref{eq:quantumadiabaticlength}) as

\begin{equation}
d\mathcal{L}_{\{n(k)\}}=\left|d\lambda\right|\left[\sum_{\{l(k)\}\ne\{n(k)\}}\left|\frac{\left\langle \Psi_{\{l(k)\}}\right|\frac{\partial}{\partial\lambda}\left|\Psi_{\{n(k)\}}\right\rangle }{\tilde{E}_{\{n(k)\}}(s)-\tilde{E}_{\{l(k)\}}(s)}\right|^{2}\right]^{1/2},
\end{equation}
where $\{n(k)\}$ and $\{l(k)\}$ are the eigenstates of the many-body
system with $n,l=\mathrm{g}$ or $\mathrm{e}$. The change of the
many-body eigenstate is

\begin{equation}
\frac{\partial}{\partial\lambda}\left|\Psi_{\{n(k)\}}\right\rangle =\sum_{k^{\prime}>0}\frac{\partial\left|n(k^{\prime})\right\rangle }{\partial\lambda}\otimes\underset{k\ne k^{\prime}}{\bigotimes}\left|n(k)\right\rangle .
\end{equation}
Therefore, non-zero product $\left\langle \Psi_{\{l(k)\}}\right|\partial_{\lambda}\left|\Psi_{\{n(k)\}}\right\rangle $
requires that the set $\{l(k)\}$ has only one element different from
$\{n(k)\}$. We write this different element as $n(k^{\prime})$ in
$\{n(k)\}$ and $\bar{n}(k^{\prime})$ in $\{l(k)\}$, where $\bar{n}$
is the opposite state of $n$. The non-adiabatic transition rate is
simplified as

\begin{equation}
\frac{\left\langle \Psi_{\{l(k)\}}\right|\frac{\partial}{\partial\lambda}\left|\Psi_{\{n(k)\}}\right\rangle }{\tilde{E}_{\{n(k)\}}(s)-\tilde{E}_{\{l(k)\}}(s)}=\frac{\left\langle \bar{n}(k^{\prime})\right|\frac{\partial}{\partial\lambda}\left|n(k^{\prime})\right\rangle }{\epsilon_{n(k^{\prime})}-\epsilon_{\bar{n}(k^{\prime})}}.
\end{equation}
The summation over $\{l(k)\}$ gives $N/2$ non-zero terms

\begin{equation}
\sum_{\{l(k)\}\ne\{n(k)\}}\left|\frac{\left\langle \Psi_{\{l(k)\}}\right|\frac{\partial}{\partial\lambda}\left|\Psi_{\{n(k)\}}\right\rangle }{\tilde{E}_{\{n(k)\}}(s)-\tilde{E}_{\{l(k)\}}(s)}\right|^{2}=\sum_{k^{\prime}>0}\left|\frac{\left\langle \bar{n}(k^{\prime})\right|\frac{\partial}{\partial\lambda}\left|n(k^{\prime})\right\rangle }{\epsilon_{n(k^{\prime})}-\epsilon_{\bar{n}(k^{\prime})}}\right|.
\end{equation}
The same result is obtained for both $n=\mathrm{g}$ and $\mathrm{e}$

\begin{equation}
\left|\frac{\left\langle \bar{n}(k^{\prime})\right|\frac{\partial}{\partial\lambda}\left|n(k^{\prime})\right\rangle }{\epsilon_{n(k^{\prime})}-\epsilon_{\bar{n}(k^{\prime})}}\right|^{2}=\frac{\sin^{2}k^{\prime}}{64J^{2}\left(\lambda^{2}-2\lambda\cos k^{\prime}+1\right)^{3}}.
\end{equation}
The optimal protocol Eq. (\ref{eq:optimalisingmodel}) is obtained
by varying the control parameter $\lambda$ with the constant velocity
of the quantum adiabatic length.

\end{widetext}

\bibliographystyle{apsrev4-1}
\bibliography{finite_time_adiabatic}

\begin{thebibliography}{62}%
\makeatletter
\providecommand \@ifxundefined [1]{%
 \@ifx{#1\undefined}
}%
\providecommand \@ifnum [1]{%
 \ifnum #1\expandafter \@firstoftwo
 \else \expandafter \@secondoftwo
 \fi
}%
\providecommand \@ifx [1]{%
 \ifx #1\expandafter \@firstoftwo
 \else \expandafter \@secondoftwo
 \fi
}%
\providecommand \natexlab [1]{#1}%
\providecommand \enquote  [1]{``#1''}%
\providecommand \bibnamefont  [1]{#1}%
\providecommand \bibfnamefont [1]{#1}%
\providecommand \citenamefont [1]{#1}%
\providecommand \href@noop [0]{\@secondoftwo}%
\providecommand \href [0]{\begingroup \@sanitize@url \@href}%
\providecommand \@href[1]{\@@startlink{#1}\@@href}%
\providecommand \@@href[1]{\endgroup#1\@@endlink}%
\providecommand \@sanitize@url [0]{\catcode `\\12\catcode `\$12\catcode
  `\&12\catcode `\#12\catcode `\^12\catcode `\_12\catcode `\%12\relax}%
\providecommand \@@startlink[1]{}%
\providecommand \@@endlink[0]{}%
\providecommand \url  [0]{\begingroup\@sanitize@url \@url }%
\providecommand \@url [1]{\endgroup\@href {#1}{\urlprefix }}%
\providecommand \urlprefix  [0]{URL }%
\providecommand \Eprint [0]{\href }%
\providecommand \doibase [0]{http://dx.doi.org/}%
\providecommand \selectlanguage [0]{\@gobble}%
\providecommand \bibinfo  [0]{\@secondoftwo}%
\providecommand \bibfield  [0]{\@secondoftwo}%
\providecommand \translation [1]{[#1]}%
\providecommand \BibitemOpen [0]{}%
\providecommand \bibitemStop [0]{}%
\providecommand \bibitemNoStop [0]{.\EOS\space}%
\providecommand \EOS [0]{\spacefactor3000\relax}%
\providecommand \BibitemShut  [1]{\csname bibitem#1\endcsname}%
\let\auto@bib@innerbib\@empty
\bibitem [{\citenamefont {Peirce}\ \emph {et~al.}(1988)\citenamefont {Peirce},
  \citenamefont {Dahleh},\ and\ \citenamefont {Rabitz}}]{Peirce1988}%
  \BibitemOpen
  \bibfield  {author} {\bibinfo {author} {\bibfnamefont {A.~P.}\ \bibnamefont
  {Peirce}}, \bibinfo {author} {\bibfnamefont {M.~A.}\ \bibnamefont {Dahleh}},
  \ and\ \bibinfo {author} {\bibfnamefont {H.}~\bibnamefont {Rabitz}},\ }\href
  {\doibase 10.1103/physreva.37.4950} {\bibfield  {journal} {\bibinfo
  {journal} {Phys. Rev. A}\ }\textbf {\bibinfo {volume} {37}},\ \bibinfo
  {pages} {4950} (\bibinfo {year} {1988})}\BibitemShut {NoStop}%
\bibitem [{\citenamefont {Caneva}\ \emph {et~al.}(2009)\citenamefont {Caneva},
  \citenamefont {Murphy}, \citenamefont {Calarco}, \citenamefont {Fazio},
  \citenamefont {Montangero}, \citenamefont {Giovannetti},\ and\ \citenamefont
  {Santoro}}]{Caneva2009}%
  \BibitemOpen
  \bibfield  {author} {\bibinfo {author} {\bibfnamefont {T.}~\bibnamefont
  {Caneva}}, \bibinfo {author} {\bibfnamefont {M.}~\bibnamefont {Murphy}},
  \bibinfo {author} {\bibfnamefont {T.}~\bibnamefont {Calarco}}, \bibinfo
  {author} {\bibfnamefont {R.}~\bibnamefont {Fazio}}, \bibinfo {author}
  {\bibfnamefont {S.}~\bibnamefont {Montangero}}, \bibinfo {author}
  {\bibfnamefont {V.}~\bibnamefont {Giovannetti}}, \ and\ \bibinfo {author}
  {\bibfnamefont {G.~E.}\ \bibnamefont {Santoro}},\ }\href {\doibase
  10.1103/physrevlett.103.240501} {\bibfield  {journal} {\bibinfo  {journal}
  {Phys. Rev. Lett.}\ }\textbf {\bibinfo {volume} {103}},\ \bibinfo {pages}
  {240501} (\bibinfo {year} {2009})}\BibitemShut {NoStop}%
\bibitem [{\citenamefont {Bason}\ \emph {et~al.}(2011)\citenamefont {Bason},
  \citenamefont {Viteau}, \citenamefont {Malossi}, \citenamefont {Huillery},
  \citenamefont {Arimondo}, \citenamefont {Ciampini}, \citenamefont {Fazio},
  \citenamefont {Giovannetti}, \citenamefont {Mannella},\ and\ \citenamefont
  {Morsch}}]{Bason2011}%
  \BibitemOpen
  \bibfield  {author} {\bibinfo {author} {\bibfnamefont {M.~G.}\ \bibnamefont
  {Bason}}, \bibinfo {author} {\bibfnamefont {M.}~\bibnamefont {Viteau}},
  \bibinfo {author} {\bibfnamefont {N.}~\bibnamefont {Malossi}}, \bibinfo
  {author} {\bibfnamefont {P.}~\bibnamefont {Huillery}}, \bibinfo {author}
  {\bibfnamefont {E.}~\bibnamefont {Arimondo}}, \bibinfo {author}
  {\bibfnamefont {D.}~\bibnamefont {Ciampini}}, \bibinfo {author}
  {\bibfnamefont {R.}~\bibnamefont {Fazio}}, \bibinfo {author} {\bibfnamefont
  {V.}~\bibnamefont {Giovannetti}}, \bibinfo {author} {\bibfnamefont
  {R.}~\bibnamefont {Mannella}}, \ and\ \bibinfo {author} {\bibfnamefont
  {O.}~\bibnamefont {Morsch}},\ }\href {\doibase 10.1038/nphys2170} {\bibfield
  {journal} {\bibinfo  {journal} {Nat. Phys.}\ }\textbf {\bibinfo {volume}
  {8}},\ \bibinfo {pages} {147} (\bibinfo {year} {2011})}\BibitemShut {NoStop}%
\bibitem [{\citenamefont {Brif}\ \emph {et~al.}(2014)\citenamefont {Brif},
  \citenamefont {Grace}, \citenamefont {Sarovar},\ and\ \citenamefont
  {Young}}]{Brif2014}%
  \BibitemOpen
  \bibfield  {author} {\bibinfo {author} {\bibfnamefont {C.}~\bibnamefont
  {Brif}}, \bibinfo {author} {\bibfnamefont {M.~D.}\ \bibnamefont {Grace}},
  \bibinfo {author} {\bibfnamefont {M.}~\bibnamefont {Sarovar}}, \ and\
  \bibinfo {author} {\bibfnamefont {K.~C.}\ \bibnamefont {Young}},\ }\href
  {\doibase 10.1088/1367-2630/16/6/065013} {\bibfield  {journal} {\bibinfo
  {journal} {New J. Phys.}\ }\textbf {\bibinfo {volume} {16}},\ \bibinfo
  {pages} {065013} (\bibinfo {year} {2014})}\BibitemShut {NoStop}%
\bibitem [{\citenamefont {Santos}\ and\ \citenamefont
  {Sarandy}(2015)}]{Santos2015}%
  \BibitemOpen
  \bibfield  {author} {\bibinfo {author} {\bibfnamefont {A.~C.}\ \bibnamefont
  {Santos}}\ and\ \bibinfo {author} {\bibfnamefont {M.~S.}\ \bibnamefont
  {Sarandy}},\ }\href {\doibase 10.1038/srep15775} {\bibfield  {journal}
  {\bibinfo  {journal} {Sci. Rep.}\ }\textbf {\bibinfo {volume} {5}},\ \bibinfo
  {pages} {15775} (\bibinfo {year} {2015})}\BibitemShut {NoStop}%
\bibitem [{\citenamefont {Machnes}\ \emph {et~al.}(2018)\citenamefont
  {Machnes}, \citenamefont {Ass{\'{e}}mat}, \citenamefont {Tannor},\ and\
  \citenamefont {Wilhelm}}]{Machnes2018}%
  \BibitemOpen
  \bibfield  {author} {\bibinfo {author} {\bibfnamefont {S.}~\bibnamefont
  {Machnes}}, \bibinfo {author} {\bibfnamefont {E.}~\bibnamefont
  {Ass{\'{e}}mat}}, \bibinfo {author} {\bibfnamefont {D.}~\bibnamefont
  {Tannor}}, \ and\ \bibinfo {author} {\bibfnamefont {F.~K.}\ \bibnamefont
  {Wilhelm}},\ }\href {\doibase 10.1103/physrevlett.120.150401} {\bibfield
  {journal} {\bibinfo  {journal} {Phys. Rev. Lett.}\ }\textbf {\bibinfo
  {volume} {120}},\ \bibinfo {pages} {150401} (\bibinfo {year}
  {2018})}\BibitemShut {NoStop}%
\bibitem [{\citenamefont {Zulkowski}\ and\ \citenamefont
  {DeWeese}(2015)}]{Zulkowski2015}%
  \BibitemOpen
  \bibfield  {author} {\bibinfo {author} {\bibfnamefont {P.~R.}\ \bibnamefont
  {Zulkowski}}\ and\ \bibinfo {author} {\bibfnamefont {M.~R.}\ \bibnamefont
  {DeWeese}},\ }\href {\doibase 10.1103/physreve.92.032113} {\bibfield
  {journal} {\bibinfo  {journal} {Phys. Rev. E}\ }\textbf {\bibinfo {volume}
  {92}},\ \bibinfo {pages} {032113} (\bibinfo {year} {2015})}\BibitemShut
  {NoStop}%
\bibitem [{\citenamefont {Solon}\ and\ \citenamefont
  {Horowitz}(2018)}]{Solon2018}%
  \BibitemOpen
  \bibfield  {author} {\bibinfo {author} {\bibfnamefont {A.~P.}\ \bibnamefont
  {Solon}}\ and\ \bibinfo {author} {\bibfnamefont {J.~M.}\ \bibnamefont
  {Horowitz}},\ }\href {\doibase 10.1103/physrevlett.120.180605} {\bibfield
  {journal} {\bibinfo  {journal} {Phys. Rev. Lett.}\ }\textbf {\bibinfo
  {volume} {120}},\ \bibinfo {pages} {180605} (\bibinfo {year}
  {2018})}\BibitemShut {NoStop}%
\bibitem [{\citenamefont {Cavina}\ \emph {et~al.}(2018)\citenamefont {Cavina},
  \citenamefont {Mari}, \citenamefont {Carlini},\ and\ \citenamefont
  {Giovannetti}}]{Cavina2018}%
  \BibitemOpen
  \bibfield  {author} {\bibinfo {author} {\bibfnamefont {V.}~\bibnamefont
  {Cavina}}, \bibinfo {author} {\bibfnamefont {A.}~\bibnamefont {Mari}},
  \bibinfo {author} {\bibfnamefont {A.}~\bibnamefont {Carlini}}, \ and\
  \bibinfo {author} {\bibfnamefont {V.}~\bibnamefont {Giovannetti}},\ }\href
  {\doibase 10.1103/physreva.98.052125} {\bibfield  {journal} {\bibinfo
  {journal} {Phys. Rev. A}\ }\textbf {\bibinfo {volume} {98}},\ \bibinfo
  {pages} {052125} (\bibinfo {year} {2018})}\BibitemShut {NoStop}%
\bibitem [{\citenamefont {Scandi}\ and\ \citenamefont
  {Perarnau-Llobet}(2019)}]{Scandi2018}%
  \BibitemOpen
  \bibfield  {author} {\bibinfo {author} {\bibfnamefont {M.}~\bibnamefont
  {Scandi}}\ and\ \bibinfo {author} {\bibfnamefont {M.}~\bibnamefont
  {Perarnau-Llobet}},\ }\href {\doibase 10.22331/q-2019-10-24-197} {\bibfield
  {journal} {\bibinfo  {journal} {Quantum}\ }\textbf {\bibinfo {volume} {3}},\
  \bibinfo {pages} {197} (\bibinfo {year} {2019})}\BibitemShut {NoStop}%
\bibitem [{\citenamefont {Vu}\ and\ \citenamefont {Saito}(2022)}]{Vu2022}%
  \BibitemOpen
  \bibfield  {author} {\bibinfo {author} {\bibfnamefont {T.~V.}\ \bibnamefont
  {Vu}}\ and\ \bibinfo {author} {\bibfnamefont {K.}~\bibnamefont {Saito}},\
  }\href {\doibase 10.1103/physrevlett.128.010602} {\bibfield  {journal}
  {\bibinfo  {journal} {Phys. Rev. Lett.}\ }\textbf {\bibinfo {volume} {128}},\
  \bibinfo {pages} {010602} (\bibinfo {year} {2022})}\BibitemShut {NoStop}%
\bibitem [{\citenamefont {Farhi}\ \emph {et~al.}(2001)\citenamefont {Farhi},
  \citenamefont {Goldstone}, \citenamefont {Gutmann}, \citenamefont {Lapan},
  \citenamefont {Lundgren},\ and\ \citenamefont {Preda}}]{Farhi2001}%
  \BibitemOpen
  \bibfield  {author} {\bibinfo {author} {\bibfnamefont {E.}~\bibnamefont
  {Farhi}}, \bibinfo {author} {\bibfnamefont {J.}~\bibnamefont {Goldstone}},
  \bibinfo {author} {\bibfnamefont {S.}~\bibnamefont {Gutmann}}, \bibinfo
  {author} {\bibfnamefont {J.}~\bibnamefont {Lapan}}, \bibinfo {author}
  {\bibfnamefont {A.}~\bibnamefont {Lundgren}}, \ and\ \bibinfo {author}
  {\bibfnamefont {D.}~\bibnamefont {Preda}},\ }\href {\doibase
  10.1126/science.1057726} {\bibfield  {journal} {\bibinfo  {journal}
  {Science}\ }\textbf {\bibinfo {volume} {292}},\ \bibinfo {pages} {472}
  (\bibinfo {year} {2001})}\BibitemShut {NoStop}%
\bibitem [{\citenamefont {Sarandy}\ and\ \citenamefont
  {Lidar}(2005)}]{Sarandy2005}%
  \BibitemOpen
  \bibfield  {author} {\bibinfo {author} {\bibfnamefont {M.~S.}\ \bibnamefont
  {Sarandy}}\ and\ \bibinfo {author} {\bibfnamefont {D.~A.}\ \bibnamefont
  {Lidar}},\ }\href {\doibase 10.1103/physrevlett.95.250503} {\bibfield
  {journal} {\bibinfo  {journal} {Phys. Rev. Lett.}\ }\textbf {\bibinfo
  {volume} {95}},\ \bibinfo {pages} {250503} (\bibinfo {year}
  {2005})}\BibitemShut {NoStop}%
\bibitem [{\citenamefont {Nielsen}(2006)}]{Nielsen2006}%
  \BibitemOpen
  \bibfield  {author} {\bibinfo {author} {\bibfnamefont {M.~A.}\ \bibnamefont
  {Nielsen}},\ }\href {\doibase 10.1126/science.1121541} {\bibfield  {journal}
  {\bibinfo  {journal} {Science}\ }\textbf {\bibinfo {volume} {311}},\ \bibinfo
  {pages} {1133} (\bibinfo {year} {2006})}\BibitemShut {NoStop}%
\bibitem [{\citenamefont {Menicucci}\ \emph {et~al.}(2006)\citenamefont
  {Menicucci}, \citenamefont {van Loock}, \citenamefont {Gu}, \citenamefont
  {Weedbrook}, \citenamefont {Ralph},\ and\ \citenamefont
  {Nielsen}}]{Menicucci2006}%
  \BibitemOpen
  \bibfield  {author} {\bibinfo {author} {\bibfnamefont {N.~C.}\ \bibnamefont
  {Menicucci}}, \bibinfo {author} {\bibfnamefont {P.}~\bibnamefont {van
  Loock}}, \bibinfo {author} {\bibfnamefont {M.}~\bibnamefont {Gu}}, \bibinfo
  {author} {\bibfnamefont {C.}~\bibnamefont {Weedbrook}}, \bibinfo {author}
  {\bibfnamefont {T.~C.}\ \bibnamefont {Ralph}}, \ and\ \bibinfo {author}
  {\bibfnamefont {M.~A.}\ \bibnamefont {Nielsen}},\ }\href {\doibase
  10.1103/physrevlett.97.110501} {\bibfield  {journal} {\bibinfo  {journal}
  {Phys. Rev. Lett.}\ }\textbf {\bibinfo {volume} {97}},\ \bibinfo {pages}
  {110501} (\bibinfo {year} {2006})}\BibitemShut {NoStop}%
\bibitem [{\citenamefont {Aharonov}\ \emph {et~al.}(2007)\citenamefont
  {Aharonov}, \citenamefont {van Dam}, \citenamefont {Kempe}, \citenamefont
  {Landau}, \citenamefont {Lloyd},\ and\ \citenamefont {Regev}}]{Aharonov2007}%
  \BibitemOpen
  \bibfield  {author} {\bibinfo {author} {\bibfnamefont {D.}~\bibnamefont
  {Aharonov}}, \bibinfo {author} {\bibfnamefont {W.}~\bibnamefont {van Dam}},
  \bibinfo {author} {\bibfnamefont {J.}~\bibnamefont {Kempe}}, \bibinfo
  {author} {\bibfnamefont {Z.}~\bibnamefont {Landau}}, \bibinfo {author}
  {\bibfnamefont {S.}~\bibnamefont {Lloyd}}, \ and\ \bibinfo {author}
  {\bibfnamefont {O.}~\bibnamefont {Regev}},\ }\href {\doibase
  10.1137/s0097539705447323} {\bibfield  {journal} {\bibinfo  {journal} {SIAM
  J. Comput.}\ }\textbf {\bibinfo {volume} {37}},\ \bibinfo {pages} {166}
  (\bibinfo {year} {2007})}\BibitemShut {NoStop}%
\bibitem [{\citenamefont {Albash}\ and\ \citenamefont
  {Lidar}(2018)}]{Albash2018}%
  \BibitemOpen
  \bibfield  {author} {\bibinfo {author} {\bibfnamefont {T.}~\bibnamefont
  {Albash}}\ and\ \bibinfo {author} {\bibfnamefont {D.~A.}\ \bibnamefont
  {Lidar}},\ }\href {\doibase 10.1103/revmodphys.90.015002} {\bibfield
  {journal} {\bibinfo  {journal} {Rev. Mod. Phys.}\ }\textbf {\bibinfo {volume}
  {90}},\ \bibinfo {pages} {015002} (\bibinfo {year} {2018})}\BibitemShut
  {NoStop}%
\bibitem [{\citenamefont {Feldmann}\ and\ \citenamefont
  {Kosloff}(2000)}]{Feldmann2000}%
  \BibitemOpen
  \bibfield  {author} {\bibinfo {author} {\bibfnamefont {T.}~\bibnamefont
  {Feldmann}}\ and\ \bibinfo {author} {\bibfnamefont {R.}~\bibnamefont
  {Kosloff}},\ }\href {\doibase 10.1103/physreve.61.4774} {\bibfield  {journal}
  {\bibinfo  {journal} {Phys. Rev. E}\ }\textbf {\bibinfo {volume} {61}},\
  \bibinfo {pages} {4774} (\bibinfo {year} {2000})}\BibitemShut {NoStop}%
\bibitem [{\citenamefont {Kieu}(2004)}]{Kieu2004}%
  \BibitemOpen
  \bibfield  {author} {\bibinfo {author} {\bibfnamefont {T.~D.}\ \bibnamefont
  {Kieu}},\ }\href {\doibase 10.1103/physrevlett.93.140403} {\bibfield
  {journal} {\bibinfo  {journal} {Phys. Rev. Lett.}\ }\textbf {\bibinfo
  {volume} {93}},\ \bibinfo {pages} {140403} (\bibinfo {year}
  {2004})}\BibitemShut {NoStop}%
\bibitem [{\citenamefont {Quan}\ \emph {et~al.}(2007)\citenamefont {Quan},
  \citenamefont {Liu}, \citenamefont {Sun},\ and\ \citenamefont
  {Nori}}]{Quan_2007}%
  \BibitemOpen
  \bibfield  {author} {\bibinfo {author} {\bibfnamefont {H.~T.}\ \bibnamefont
  {Quan}}, \bibinfo {author} {\bibfnamefont {Y.~X.}\ \bibnamefont {Liu}},
  \bibinfo {author} {\bibfnamefont {C.~P.}\ \bibnamefont {Sun}}, \ and\
  \bibinfo {author} {\bibfnamefont {F.}~\bibnamefont {Nori}},\ }\href {\doibase
  10.1103/physreve.76.031105} {\bibfield  {journal} {\bibinfo  {journal} {Phys.
  Rev. E}\ }\textbf {\bibinfo {volume} {76}},\ \bibinfo {pages} {031105}
  (\bibinfo {year} {2007})}\BibitemShut {NoStop}%
\bibitem [{\citenamefont {Quan}(2009)}]{Quan2009}%
  \BibitemOpen
  \bibfield  {author} {\bibinfo {author} {\bibfnamefont {H.~T.}\ \bibnamefont
  {Quan}},\ }\href {\doibase 10.1103/physreve.79.041129} {\bibfield  {journal}
  {\bibinfo  {journal} {Phys. Rev. E}\ }\textbf {\bibinfo {volume} {79}},\
  \bibinfo {pages} {041129} (\bibinfo {year} {2009})}\BibitemShut {NoStop}%
\bibitem [{\citenamefont {Plastina}\ \emph {et~al.}(2014)\citenamefont
  {Plastina}, \citenamefont {Alecce}, \citenamefont {Apollaro}, \citenamefont
  {Falcone}, \citenamefont {Francica}, \citenamefont {Galve}, \citenamefont
  {Gullo},\ and\ \citenamefont {Zambrini}}]{Plastina2014}%
  \BibitemOpen
  \bibfield  {author} {\bibinfo {author} {\bibfnamefont {F.}~\bibnamefont
  {Plastina}}, \bibinfo {author} {\bibfnamefont {A.}~\bibnamefont {Alecce}},
  \bibinfo {author} {\bibfnamefont {T.}~\bibnamefont {Apollaro}}, \bibinfo
  {author} {\bibfnamefont {G.}~\bibnamefont {Falcone}}, \bibinfo {author}
  {\bibfnamefont {G.}~\bibnamefont {Francica}}, \bibinfo {author}
  {\bibfnamefont {F.}~\bibnamefont {Galve}}, \bibinfo {author} {\bibfnamefont
  {N.~L.}\ \bibnamefont {Gullo}}, \ and\ \bibinfo {author} {\bibfnamefont
  {R.}~\bibnamefont {Zambrini}},\ }\href {\doibase
  10.1103/physrevlett.113.260601} {\bibfield  {journal} {\bibinfo  {journal}
  {Phys. Rev. Lett.}\ }\textbf {\bibinfo {volume} {113}},\ \bibinfo {pages}
  {260601} (\bibinfo {year} {2014})}\BibitemShut {NoStop}%
\bibitem [{\citenamefont {Amin}(2009)}]{Amin2009}%
  \BibitemOpen
  \bibfield  {author} {\bibinfo {author} {\bibfnamefont {M.~H.~S.}\
  \bibnamefont {Amin}},\ }\href {\doibase 10.1103/physrevlett.102.220401}
  {\bibfield  {journal} {\bibinfo  {journal} {Phys. Rev. Lett.}\ }\textbf
  {\bibinfo {volume} {102}},\ \bibinfo {pages} {220401} (\bibinfo {year}
  {2009})}\BibitemShut {NoStop}%
\bibitem [{\citenamefont {Sakurai}(2011)}]{Sakurai2011}%
  \BibitemOpen
  \bibfield  {author} {\bibinfo {author} {\bibfnamefont {J.~J.}\ \bibnamefont
  {Sakurai}},\ }\href@noop {} {\emph {\bibinfo {title} {Modern Quantum
  Mechanics}}}\ (\bibinfo  {publisher} {Addison-Wesley},\ \bibinfo {address}
  {Boston},\ \bibinfo {year} {2011})\BibitemShut {NoStop}%
\bibitem [{\citenamefont {Shevchenko}\ \emph {et~al.}(2010)\citenamefont
  {Shevchenko}, \citenamefont {Ashhab},\ and\ \citenamefont
  {Nori}}]{Shevchenko2010}%
  \BibitemOpen
  \bibfield  {author} {\bibinfo {author} {\bibfnamefont {S.}~\bibnamefont
  {Shevchenko}}, \bibinfo {author} {\bibfnamefont {S.}~\bibnamefont {Ashhab}},
  \ and\ \bibinfo {author} {\bibfnamefont {F.}~\bibnamefont {Nori}},\ }\href
  {\doibase 10.1016/j.physrep.2010.03.002} {\bibfield  {journal} {\bibinfo
  {journal} {Phys. Rep.}\ }\textbf {\bibinfo {volume} {492}},\ \bibinfo {pages}
  {1} (\bibinfo {year} {2010})}\BibitemShut {NoStop}%
\bibitem [{\citenamefont {Demirplak}\ and\ \citenamefont
  {Rice}(2003)}]{Demirplak2003}%
  \BibitemOpen
  \bibfield  {author} {\bibinfo {author} {\bibfnamefont {M.}~\bibnamefont
  {Demirplak}}\ and\ \bibinfo {author} {\bibfnamefont {S.~A.}\ \bibnamefont
  {Rice}},\ }\href {\doibase 10.1021/jp030708a} {\bibfield  {journal} {\bibinfo
   {journal} {J. Phys. Chem. A}\ }\textbf {\bibinfo {volume} {107}},\ \bibinfo
  {pages} {9937} (\bibinfo {year} {2003})}\BibitemShut {NoStop}%
\bibitem [{\citenamefont {Demirplak}\ and\ \citenamefont
  {Rice}(2005)}]{Demirplak2005}%
  \BibitemOpen
  \bibfield  {author} {\bibinfo {author} {\bibfnamefont {M.}~\bibnamefont
  {Demirplak}}\ and\ \bibinfo {author} {\bibfnamefont {S.~A.}\ \bibnamefont
  {Rice}},\ }\href {\doibase 10.1021/jp040647w} {\bibfield  {journal} {\bibinfo
   {journal} {J. Phys. Chem. B}\ }\textbf {\bibinfo {volume} {109}},\ \bibinfo
  {pages} {6838} (\bibinfo {year} {2005})}\BibitemShut {NoStop}%
\bibitem [{\citenamefont {Masuda}\ and\ \citenamefont
  {Nakamura}(2008)}]{Masuda2008}%
  \BibitemOpen
  \bibfield  {author} {\bibinfo {author} {\bibfnamefont {S.}~\bibnamefont
  {Masuda}}\ and\ \bibinfo {author} {\bibfnamefont {K.}~\bibnamefont
  {Nakamura}},\ }\href {\doibase 10.1103/physreva.78.062108} {\bibfield
  {journal} {\bibinfo  {journal} {Phys. Rev. A}\ }\textbf {\bibinfo {volume}
  {78}},\ \bibinfo {pages} {062108} (\bibinfo {year} {2008})}\BibitemShut
  {NoStop}%
\bibitem [{\citenamefont {Berry}(2009)}]{Berry2009}%
  \BibitemOpen
  \bibfield  {author} {\bibinfo {author} {\bibfnamefont {M.~V.}\ \bibnamefont
  {Berry}},\ }\href {\doibase 10.1088/1751-8113/42/36/365303} {\bibfield
  {journal} {\bibinfo  {journal} {J. Phys. A: Math. Theor.}\ }\textbf {\bibinfo
  {volume} {42}},\ \bibinfo {pages} {365303} (\bibinfo {year}
  {2009})}\BibitemShut {NoStop}%
\bibitem [{\citenamefont {Chen}\ \emph {et~al.}(2010)\citenamefont {Chen},
  \citenamefont {Ruschhaupt}, \citenamefont {Schmidt}, \citenamefont {del
  Campo}, \citenamefont {Gu{\'{e}}ry-Odelin},\ and\ \citenamefont
  {Muga}}]{Chen2010PhysRevLett104_63002}%
  \BibitemOpen
  \bibfield  {author} {\bibinfo {author} {\bibfnamefont {X.}~\bibnamefont
  {Chen}}, \bibinfo {author} {\bibfnamefont {A.}~\bibnamefont {Ruschhaupt}},
  \bibinfo {author} {\bibfnamefont {S.}~\bibnamefont {Schmidt}}, \bibinfo
  {author} {\bibfnamefont {A.}~\bibnamefont {del Campo}}, \bibinfo {author}
  {\bibfnamefont {D.}~\bibnamefont {Gu{\'{e}}ry-Odelin}}, \ and\ \bibinfo
  {author} {\bibfnamefont {J.~G.}\ \bibnamefont {Muga}},\ }\href {\doibase
  10.1103/physrevlett.104.063002} {\bibfield  {journal} {\bibinfo  {journal}
  {Phys. Rev. Lett.}\ }\textbf {\bibinfo {volume} {104}},\ \bibinfo {pages}
  {063002} (\bibinfo {year} {2010})}\BibitemShut {NoStop}%
\bibitem [{\citenamefont {del Campo}(2013)}]{Campo2013}%
  \BibitemOpen
  \bibfield  {author} {\bibinfo {author} {\bibfnamefont {A.}~\bibnamefont {del
  Campo}},\ }\href {\doibase 10.1103/physrevlett.111.100502} {\bibfield
  {journal} {\bibinfo  {journal} {Phys. Rev. Lett.}\ }\textbf {\bibinfo
  {volume} {111}},\ \bibinfo {pages} {100502} (\bibinfo {year}
  {2013})}\BibitemShut {NoStop}%
\bibitem [{\citenamefont {Santos}\ and\ \citenamefont
  {Sarandy}(2017)}]{Santos2017}%
  \BibitemOpen
  \bibfield  {author} {\bibinfo {author} {\bibfnamefont {A.~C.}\ \bibnamefont
  {Santos}}\ and\ \bibinfo {author} {\bibfnamefont {M.~S.}\ \bibnamefont
  {Sarandy}},\ }\href {\doibase 10.1088/1751-8121/aa96f1} {\bibfield  {journal}
  {\bibinfo  {journal} {J. Phys. A: Math. Theor.}\ }\textbf {\bibinfo {volume}
  {51}},\ \bibinfo {pages} {025301} (\bibinfo {year} {2017})}\BibitemShut
  {NoStop}%
\bibitem [{\citenamefont {Gu{\'{e}}ry-Odelin}\ \emph
  {et~al.}(2019)\citenamefont {Gu{\'{e}}ry-Odelin}, \citenamefont {Ruschhaupt},
  \citenamefont {Kiely}, \citenamefont {Torrontegui}, \citenamefont
  {Mart{\'{\i}}nez-Garaot},\ and\ \citenamefont {Muga}}]{Guery-Odelin2019}%
  \BibitemOpen
  \bibfield  {author} {\bibinfo {author} {\bibfnamefont {D.}~\bibnamefont
  {Gu{\'{e}}ry-Odelin}}, \bibinfo {author} {\bibfnamefont {A.}~\bibnamefont
  {Ruschhaupt}}, \bibinfo {author} {\bibfnamefont {A.}~\bibnamefont {Kiely}},
  \bibinfo {author} {\bibfnamefont {E.}~\bibnamefont {Torrontegui}}, \bibinfo
  {author} {\bibfnamefont {S.}~\bibnamefont {Mart{\'{\i}}nez-Garaot}}, \ and\
  \bibinfo {author} {\bibfnamefont {J.}~\bibnamefont {Muga}},\ }\href {\doibase
  10.1103/revmodphys.91.045001} {\bibfield  {journal} {\bibinfo  {journal}
  {Rev. Mod. Phys.}\ }\textbf {\bibinfo {volume} {91}},\ \bibinfo {pages}
  {045001} (\bibinfo {year} {2019})}\BibitemShut {NoStop}%
\bibitem [{\citenamefont {Hu}\ \emph {et~al.}(2018)\citenamefont {Hu},
  \citenamefont {Cui}, \citenamefont {Santos}, \citenamefont {Huang},
  \citenamefont {Sarandy}, \citenamefont {Li},\ and\ \citenamefont
  {Guo}}]{Hu2018}%
  \BibitemOpen
  \bibfield  {author} {\bibinfo {author} {\bibfnamefont {C.-K.}\ \bibnamefont
  {Hu}}, \bibinfo {author} {\bibfnamefont {J.-M.}\ \bibnamefont {Cui}},
  \bibinfo {author} {\bibfnamefont {A.~C.}\ \bibnamefont {Santos}}, \bibinfo
  {author} {\bibfnamefont {Y.-F.}\ \bibnamefont {Huang}}, \bibinfo {author}
  {\bibfnamefont {M.~S.}\ \bibnamefont {Sarandy}}, \bibinfo {author}
  {\bibfnamefont {C.-F.}\ \bibnamefont {Li}}, \ and\ \bibinfo {author}
  {\bibfnamefont {G.-C.}\ \bibnamefont {Guo}},\ }\href {\doibase
  10.1364/ol.43.003136} {\bibfield  {journal} {\bibinfo  {journal} {Opt.
  Lett.}\ }\textbf {\bibinfo {volume} {43}},\ \bibinfo {pages} {3136} (\bibinfo
  {year} {2018})}\BibitemShut {NoStop}%
\bibitem [{\citenamefont {Mart{\'{\i}}nez-Garaot}\ \emph
  {et~al.}(2015)\citenamefont {Mart{\'{\i}}nez-Garaot}, \citenamefont
  {Ruschhaupt}, \citenamefont {Gillet}, \citenamefont {Busch},\ and\
  \citenamefont {Muga}}]{Martinez-Garaot2015}%
  \BibitemOpen
  \bibfield  {author} {\bibinfo {author} {\bibfnamefont {S.}~\bibnamefont
  {Mart{\'{\i}}nez-Garaot}}, \bibinfo {author} {\bibfnamefont {A.}~\bibnamefont
  {Ruschhaupt}}, \bibinfo {author} {\bibfnamefont {J.}~\bibnamefont {Gillet}},
  \bibinfo {author} {\bibfnamefont {T.}~\bibnamefont {Busch}}, \ and\ \bibinfo
  {author} {\bibfnamefont {J.~G.}\ \bibnamefont {Muga}},\ }\href {\doibase
  10.1103/physreva.92.043406} {\bibfield  {journal} {\bibinfo  {journal} {Phys.
  Rev. A}\ }\textbf {\bibinfo {volume} {92}},\ \bibinfo {pages} {043406}
  (\bibinfo {year} {2015})}\BibitemShut {NoStop}%
\bibitem [{\citenamefont {Chung}\ \emph {et~al.}(2017)\citenamefont {Chung},
  \citenamefont {Lee},\ and\ \citenamefont {Tseng}}]{Chung2017}%
  \BibitemOpen
  \bibfield  {author} {\bibinfo {author} {\bibfnamefont {H.-C.}\ \bibnamefont
  {Chung}}, \bibinfo {author} {\bibfnamefont {K.-S.}\ \bibnamefont {Lee}}, \
  and\ \bibinfo {author} {\bibfnamefont {S.-Y.}\ \bibnamefont {Tseng}},\ }\href
  {\doibase 10.1364/oe.25.013626} {\bibfield  {journal} {\bibinfo  {journal}
  {Opt. Express}\ }\textbf {\bibinfo {volume} {25}},\ \bibinfo {pages} {13626}
  (\bibinfo {year} {2017})}\BibitemShut {NoStop}%
\bibitem [{\citenamefont {Mart{\'{\i}}nez-Garaot}\ \emph
  {et~al.}(2017)\citenamefont {Mart{\'{\i}}nez-Garaot}, \citenamefont {Muga},\
  and\ \citenamefont {Tseng}}]{MartinezGaraot2017}%
  \BibitemOpen
  \bibfield  {author} {\bibinfo {author} {\bibfnamefont {S.}~\bibnamefont
  {Mart{\'{\i}}nez-Garaot}}, \bibinfo {author} {\bibfnamefont {J.~G.}\
  \bibnamefont {Muga}}, \ and\ \bibinfo {author} {\bibfnamefont {S.-Y.}\
  \bibnamefont {Tseng}},\ }\href {\doibase 10.1364/oe.25.000159} {\bibfield
  {journal} {\bibinfo  {journal} {Opt. Express}\ }\textbf {\bibinfo {volume}
  {25}},\ \bibinfo {pages} {159} (\bibinfo {year} {2017})}\BibitemShut
  {NoStop}%
\bibitem [{\citenamefont {Liu}\ and\ \citenamefont {Tseng}(2017)}]{Liu2017}%
  \BibitemOpen
  \bibfield  {author} {\bibinfo {author} {\bibfnamefont {Y.-H.}\ \bibnamefont
  {Liu}}\ and\ \bibinfo {author} {\bibfnamefont {S.-Y.}\ \bibnamefont
  {Tseng}},\ }\href {\doibase 10.1088/1361-6455/aa8b4e} {\bibfield  {journal}
  {\bibinfo  {journal} {J. Phys. B: At., Mol. Opt. Phys.}\ }\textbf {\bibinfo
  {volume} {50}},\ \bibinfo {pages} {205501} (\bibinfo {year}
  {2017})}\BibitemShut {NoStop}%
\bibitem [{\citenamefont {Sun}(1988)}]{Sun_1988}%
  \BibitemOpen
  \bibfield  {author} {\bibinfo {author} {\bibfnamefont {C.-P.}\ \bibnamefont
  {Sun}},\ }\href {\doibase 10.1088/0305-4470/21/7/023} {\bibfield  {journal}
  {\bibinfo  {journal} {J. Phys. A: Math. Gen.}\ }\textbf {\bibinfo {volume}
  {21}},\ \bibinfo {pages} {1595} (\bibinfo {year} {1988})}\BibitemShut
  {NoStop}%
\bibitem [{\citenamefont {Rigolin}\ \emph {et~al.}(2008)\citenamefont
  {Rigolin}, \citenamefont {Ortiz},\ and\ \citenamefont
  {Ponce}}]{Rigolin_2008}%
  \BibitemOpen
  \bibfield  {author} {\bibinfo {author} {\bibfnamefont {G.}~\bibnamefont
  {Rigolin}}, \bibinfo {author} {\bibfnamefont {G.}~\bibnamefont {Ortiz}}, \
  and\ \bibinfo {author} {\bibfnamefont {V.~H.}\ \bibnamefont {Ponce}},\ }\href
  {\doibase 10.1103/physreva.78.052508} {\bibfield  {journal} {\bibinfo
  {journal} {Phys. Rev. A}\ }\textbf {\bibinfo {volume} {78}},\ \bibinfo
  {pages} {052508} (\bibinfo {year} {2008})}\BibitemShut {NoStop}%
\bibitem [{\citenamefont {Chen}\ \emph
  {et~al.}(2019{\natexlab{a}})\citenamefont {Chen}, \citenamefont {Sun},\ and\
  \citenamefont {Dong}}]{Chen2019}%
  \BibitemOpen
  \bibfield  {author} {\bibinfo {author} {\bibfnamefont {J.-F.}\ \bibnamefont
  {Chen}}, \bibinfo {author} {\bibfnamefont {C.-P.}\ \bibnamefont {Sun}}, \
  and\ \bibinfo {author} {\bibfnamefont {H.}~\bibnamefont {Dong}},\ }\href
  {\doibase 10.1103/physreve.100.062140} {\bibfield  {journal} {\bibinfo
  {journal} {Phys. Rev. E}\ }\textbf {\bibinfo {volume} {100}},\ \bibinfo
  {pages} {062140} (\bibinfo {year} {2019}{\natexlab{a}})}\BibitemShut
  {NoStop}%
\bibitem [{\citenamefont {Chen}\ \emph
  {et~al.}(2019{\natexlab{b}})\citenamefont {Chen}, \citenamefont {Sun},\ and\
  \citenamefont {Dong}}]{Chen2019a}%
  \BibitemOpen
  \bibfield  {author} {\bibinfo {author} {\bibfnamefont {J.-F.}\ \bibnamefont
  {Chen}}, \bibinfo {author} {\bibfnamefont {C.-P.}\ \bibnamefont {Sun}}, \
  and\ \bibinfo {author} {\bibfnamefont {H.}~\bibnamefont {Dong}},\ }\href
  {\doibase 10.1103/physreve.100.032144} {\bibfield  {journal} {\bibinfo
  {journal} {Phys. Rev. E}\ }\textbf {\bibinfo {volume} {100}},\ \bibinfo
  {pages} {032144} (\bibinfo {year} {2019}{\natexlab{b}})}\BibitemShut
  {NoStop}%
\bibitem [{\citenamefont {Provost}\ and\ \citenamefont
  {Vallee}(1980)}]{Provost1980}%
  \BibitemOpen
  \bibfield  {author} {\bibinfo {author} {\bibfnamefont {J.~P.}\ \bibnamefont
  {Provost}}\ and\ \bibinfo {author} {\bibfnamefont {G.}~\bibnamefont
  {Vallee}},\ }\href {\doibase 10.1007/bf02193559} {\bibfield  {journal}
  {\bibinfo  {journal} {Commun. Math. Phys.}\ }\textbf {\bibinfo {volume}
  {76}},\ \bibinfo {pages} {289} (\bibinfo {year} {1980})}\BibitemShut
  {NoStop}%
\bibitem [{\citenamefont {Bengtsson}(2006)}]{Bengtsson2006}%
  \BibitemOpen
  \bibfield  {author} {\bibinfo {author} {\bibfnamefont {I.}~\bibnamefont
  {Bengtsson}},\ }\href@noop {} {\emph {\bibinfo {title} {Geometry of quantum
  states : an introduction to quantum entanglement}}}\ (\bibinfo  {publisher}
  {Cambridge University Press},\ \bibinfo {address} {Cambridge New York},\
  \bibinfo {year} {2006})\BibitemShut {NoStop}%
\bibitem [{\citenamefont {Zanardi}\ \emph {et~al.}(2007)\citenamefont
  {Zanardi}, \citenamefont {Giorda},\ and\ \citenamefont
  {Cozzini}}]{Zanardi2007}%
  \BibitemOpen
  \bibfield  {author} {\bibinfo {author} {\bibfnamefont {P.}~\bibnamefont
  {Zanardi}}, \bibinfo {author} {\bibfnamefont {P.}~\bibnamefont {Giorda}}, \
  and\ \bibinfo {author} {\bibfnamefont {M.}~\bibnamefont {Cozzini}},\ }\href
  {\doibase 10.1103/physrevlett.99.100603} {\bibfield  {journal} {\bibinfo
  {journal} {Phys. Rev. Lett.}\ }\textbf {\bibinfo {volume} {99}},\ \bibinfo
  {pages} {100603} (\bibinfo {year} {2007})}\BibitemShut {NoStop}%
\bibitem [{\citenamefont {Venuti}\ and\ \citenamefont
  {Zanardi}(2007)}]{Venuti2007}%
  \BibitemOpen
  \bibfield  {author} {\bibinfo {author} {\bibfnamefont {L.~C.}\ \bibnamefont
  {Venuti}}\ and\ \bibinfo {author} {\bibfnamefont {P.}~\bibnamefont
  {Zanardi}},\ }\href {\doibase 10.1103/physrevlett.99.095701} {\bibfield
  {journal} {\bibinfo  {journal} {Phys. Rev. Lett.}\ }\textbf {\bibinfo
  {volume} {99}},\ \bibinfo {pages} {095701} (\bibinfo {year}
  {2007})}\BibitemShut {NoStop}%
\bibitem [{\citenamefont {Rezakhani}\ \emph {et~al.}(2009)\citenamefont
  {Rezakhani}, \citenamefont {Kuo}, \citenamefont {Hamma}, \citenamefont
  {Lidar},\ and\ \citenamefont {Zanardi}}]{Rezakhani2009}%
  \BibitemOpen
  \bibfield  {author} {\bibinfo {author} {\bibfnamefont {A.~T.}\ \bibnamefont
  {Rezakhani}}, \bibinfo {author} {\bibfnamefont {W.-J.}\ \bibnamefont {Kuo}},
  \bibinfo {author} {\bibfnamefont {A.}~\bibnamefont {Hamma}}, \bibinfo
  {author} {\bibfnamefont {D.~A.}\ \bibnamefont {Lidar}}, \ and\ \bibinfo
  {author} {\bibfnamefont {P.}~\bibnamefont {Zanardi}},\ }\href {\doibase
  10.1103/physrevlett.103.080502} {\bibfield  {journal} {\bibinfo  {journal}
  {Phys. Rev. Lett.}\ }\textbf {\bibinfo {volume} {103}},\ \bibinfo {pages}
  {080502} (\bibinfo {year} {2009})}\BibitemShut {NoStop}%
\bibitem [{\citenamefont {Rezakhani}\ \emph {et~al.}(2010)\citenamefont
  {Rezakhani}, \citenamefont {Abasto}, \citenamefont {Lidar},\ and\
  \citenamefont {Zanardi}}]{Rezakhani2010a}%
  \BibitemOpen
  \bibfield  {author} {\bibinfo {author} {\bibfnamefont {A.~T.}\ \bibnamefont
  {Rezakhani}}, \bibinfo {author} {\bibfnamefont {D.~F.}\ \bibnamefont
  {Abasto}}, \bibinfo {author} {\bibfnamefont {D.~A.}\ \bibnamefont {Lidar}}, \
  and\ \bibinfo {author} {\bibfnamefont {P.}~\bibnamefont {Zanardi}},\ }\href
  {\doibase 10.1103/physreva.82.012321} {\bibfield  {journal} {\bibinfo
  {journal} {Phys. Rev. A}\ }\textbf {\bibinfo {volume} {82}},\ \bibinfo
  {pages} {012321} (\bibinfo {year} {2010})}\BibitemShut {NoStop}%
\bibitem [{\citenamefont {Deng}\ \emph {et~al.}(2015)\citenamefont {Deng},
  \citenamefont {Diao}, \citenamefont {Yu},\ and\ \citenamefont
  {Wu}}]{Deng2015}%
  \BibitemOpen
  \bibfield  {author} {\bibinfo {author} {\bibfnamefont {S.-J.}\ \bibnamefont
  {Deng}}, \bibinfo {author} {\bibfnamefont {P.-P.}\ \bibnamefont {Diao}},
  \bibinfo {author} {\bibfnamefont {Q.-L.}\ \bibnamefont {Yu}}, \ and\ \bibinfo
  {author} {\bibfnamefont {H.-B.}\ \bibnamefont {Wu}},\ }\href {\doibase
  10.1088/0256-307x/32/5/053401} {\bibfield  {journal} {\bibinfo  {journal}
  {Chin. Phys. Lett.}\ }\textbf {\bibinfo {volume} {32}},\ \bibinfo {pages}
  {053401} (\bibinfo {year} {2015})}\BibitemShut {NoStop}%
\bibitem [{\citenamefont {Deng}\ \emph {et~al.}(2018)\citenamefont {Deng},
  \citenamefont {Chenu}, \citenamefont {Diao}, \citenamefont {Li},
  \citenamefont {Yu}, \citenamefont {Coulamy}, \citenamefont {del Campo},\ and\
  \citenamefont {Wu}}]{Deng2018SciAdv4_5909}%
  \BibitemOpen
  \bibfield  {author} {\bibinfo {author} {\bibfnamefont {S.}~\bibnamefont
  {Deng}}, \bibinfo {author} {\bibfnamefont {A.}~\bibnamefont {Chenu}},
  \bibinfo {author} {\bibfnamefont {P.}~\bibnamefont {Diao}}, \bibinfo {author}
  {\bibfnamefont {F.}~\bibnamefont {Li}}, \bibinfo {author} {\bibfnamefont
  {S.}~\bibnamefont {Yu}}, \bibinfo {author} {\bibfnamefont {I.}~\bibnamefont
  {Coulamy}}, \bibinfo {author} {\bibfnamefont {A.}~\bibnamefont {del Campo}},
  \ and\ \bibinfo {author} {\bibfnamefont {H.}~\bibnamefont {Wu}},\ }\href
  {\doibase 10.1126/sciadv.aar5909} {\bibfield  {journal} {\bibinfo  {journal}
  {Sci. Adv.}\ }\textbf {\bibinfo {volume} {4}},\ \bibinfo {pages} {eaar5909}
  (\bibinfo {year} {2018})}\BibitemShut {NoStop}%
\bibitem [{\citenamefont {Zener}(1932)}]{1932a}%
  \BibitemOpen
  \bibfield  {author} {\bibinfo {author} {\bibfnamefont {C.}~\bibnamefont
  {Zener}},\ }\href {\doibase 10.1098/rspa.1932.0165} {\bibfield  {journal}
  {\bibinfo  {journal} {Proc. R. Soc. London, Ser. A}\ }\textbf {\bibinfo
  {volume} {137}},\ \bibinfo {pages} {696} (\bibinfo {year}
  {1932})}\BibitemShut {NoStop}%
\bibitem [{\citenamefont {Landau}(1932)}]{Landau1932}%
  \BibitemOpen
  \bibfield  {author} {\bibinfo {author} {\bibfnamefont {L.}~\bibnamefont
  {Landau}},\ }\href@noop {} {\bibfield  {journal} {\bibinfo  {journal} {Phys.
  Z. Sowjetunion}\ }\textbf {\bibinfo {volume} {2}},\ \bibinfo {pages} {46}
  (\bibinfo {year} {1932})}\BibitemShut {NoStop}%
\bibitem [{\citenamefont {Mullen}\ \emph {et~al.}(1989)\citenamefont {Mullen},
  \citenamefont {Ben-Jacob}, \citenamefont {Gefen},\ and\ \citenamefont
  {Schuss}}]{Mullen1989}%
  \BibitemOpen
  \bibfield  {author} {\bibinfo {author} {\bibfnamefont {K.}~\bibnamefont
  {Mullen}}, \bibinfo {author} {\bibfnamefont {E.}~\bibnamefont {Ben-Jacob}},
  \bibinfo {author} {\bibfnamefont {Y.}~\bibnamefont {Gefen}}, \ and\ \bibinfo
  {author} {\bibfnamefont {Z.}~\bibnamefont {Schuss}},\ }\href {\doibase
  10.1103/physrevlett.62.2543} {\bibfield  {journal} {\bibinfo  {journal}
  {Phys. Rev. Lett.}\ }\textbf {\bibinfo {volume} {62}},\ \bibinfo {pages}
  {2543} (\bibinfo {year} {1989})}\BibitemShut {NoStop}%
\bibitem [{\citenamefont {Yan}\ and\ \citenamefont {Wu}(2010)}]{Yan2010}%
  \BibitemOpen
  \bibfield  {author} {\bibinfo {author} {\bibfnamefont {Y.}~\bibnamefont
  {Yan}}\ and\ \bibinfo {author} {\bibfnamefont {B.}~\bibnamefont {Wu}},\
  }\href {\doibase 10.1103/physreva.81.022126} {\bibfield  {journal} {\bibinfo
  {journal} {Phys. Rev. A}\ }\textbf {\bibinfo {volume} {81}},\ \bibinfo
  {pages} {022126} (\bibinfo {year} {2010})}\BibitemShut {NoStop}%
\bibitem [{\citenamefont {Zurek}\ \emph {et~al.}(2005)\citenamefont {Zurek},
  \citenamefont {Dorner},\ and\ \citenamefont {Zoller}}]{Zurek2005}%
  \BibitemOpen
  \bibfield  {author} {\bibinfo {author} {\bibfnamefont {W.~H.}\ \bibnamefont
  {Zurek}}, \bibinfo {author} {\bibfnamefont {U.}~\bibnamefont {Dorner}}, \
  and\ \bibinfo {author} {\bibfnamefont {P.}~\bibnamefont {Zoller}},\ }\href
  {\doibase 10.1103/physrevlett.95.105701} {\bibfield  {journal} {\bibinfo
  {journal} {Phys. Rev. Lett.}\ }\textbf {\bibinfo {volume} {95}},\ \bibinfo
  {pages} {105701} (\bibinfo {year} {2005})}\BibitemShut {NoStop}%
\bibitem [{\citenamefont {Dziarmaga}(2005)}]{Dziarmaga2005}%
  \BibitemOpen
  \bibfield  {author} {\bibinfo {author} {\bibfnamefont {J.}~\bibnamefont
  {Dziarmaga}},\ }\href {\doibase 10.1103/physrevlett.95.245701} {\bibfield
  {journal} {\bibinfo  {journal} {Phys. Rev. Lett.}\ }\textbf {\bibinfo
  {volume} {95}},\ \bibinfo {pages} {245701} (\bibinfo {year}
  {2005})}\BibitemShut {NoStop}%
\bibitem [{\citenamefont {Quan}\ \emph {et~al.}(2006)\citenamefont {Quan},
  \citenamefont {Song}, \citenamefont {Liu}, \citenamefont {Zanardi},\ and\
  \citenamefont {Sun}}]{Quan2006}%
  \BibitemOpen
  \bibfield  {author} {\bibinfo {author} {\bibfnamefont {H.~T.}\ \bibnamefont
  {Quan}}, \bibinfo {author} {\bibfnamefont {Z.}~\bibnamefont {Song}}, \bibinfo
  {author} {\bibfnamefont {X.~F.}\ \bibnamefont {Liu}}, \bibinfo {author}
  {\bibfnamefont {P.}~\bibnamefont {Zanardi}}, \ and\ \bibinfo {author}
  {\bibfnamefont {C.~P.}\ \bibnamefont {Sun}},\ }\href {\doibase
  10.1103/physrevlett.96.140604} {\bibfield  {journal} {\bibinfo  {journal}
  {Phys. Rev. Lett.}\ }\textbf {\bibinfo {volume} {96}},\ \bibinfo {pages}
  {140604} (\bibinfo {year} {2006})}\BibitemShut {NoStop}%
\bibitem [{\citenamefont {Silva}(2008)}]{Silva2008}%
  \BibitemOpen
  \bibfield  {author} {\bibinfo {author} {\bibfnamefont {A.}~\bibnamefont
  {Silva}},\ }\href {\doibase 10.1103/physrevlett.101.120603} {\bibfield
  {journal} {\bibinfo  {journal} {Phys. Rev. Lett.}\ }\textbf {\bibinfo
  {volume} {101}},\ \bibinfo {pages} {120603} (\bibinfo {year}
  {2008})}\BibitemShut {NoStop}%
\bibitem [{\citenamefont {Sachdev}(2017)}]{Sachdev2017}%
  \BibitemOpen
  \bibfield  {author} {\bibinfo {author} {\bibfnamefont {S.}~\bibnamefont
  {Sachdev}},\ }\href
  {https://www.ebook.de/de/product/13782982/subir_sachdev_quantum_phase_transitions.html}
  {\emph {\bibinfo {title} {Quantum Phase Transitions}}}\ (\bibinfo
  {publisher} {Cambridge University Press},\ \bibinfo {year}
  {2017})\BibitemShut {NoStop}%
\bibitem [{\citenamefont {del Campo}(2018)}]{Campo2018a}%
  \BibitemOpen
  \bibfield  {author} {\bibinfo {author} {\bibfnamefont {A.}~\bibnamefont {del
  Campo}},\ }\href {\doibase 10.1103/physrevlett.121.200601} {\bibfield
  {journal} {\bibinfo  {journal} {Phys. Rev. Lett.}\ }\textbf {\bibinfo
  {volume} {121}},\ \bibinfo {pages} {200601} (\bibinfo {year}
  {2018})}\BibitemShut {NoStop}%
\bibitem [{\citenamefont {Fei}\ \emph {et~al.}(2020)\citenamefont {Fei},
  \citenamefont {Freitas}, \citenamefont {Cavina}, \citenamefont {Quan},\ and\
  \citenamefont {Esposito}}]{Fei2020}%
  \BibitemOpen
  \bibfield  {author} {\bibinfo {author} {\bibfnamefont {Z.}~\bibnamefont
  {Fei}}, \bibinfo {author} {\bibfnamefont {N.}~\bibnamefont {Freitas}},
  \bibinfo {author} {\bibfnamefont {V.}~\bibnamefont {Cavina}}, \bibinfo
  {author} {\bibfnamefont {H.}~\bibnamefont {Quan}}, \ and\ \bibinfo {author}
  {\bibfnamefont {M.}~\bibnamefont {Esposito}},\ }\href {\doibase
  10.1103/physrevlett.124.170603} {\bibfield  {journal} {\bibinfo  {journal}
  {Phys. Rev. Lett.}\ }\textbf {\bibinfo {volume} {124}},\ \bibinfo {pages}
  {170603} (\bibinfo {year} {2020})}\BibitemShut {NoStop}%
\bibitem [{\citenamefont {Zhang}\ and\ \citenamefont {Quan}(2022)}]{Zhang2022}%
  \BibitemOpen
  \bibfield  {author} {\bibinfo {author} {\bibfnamefont {F.}~\bibnamefont
  {Zhang}}\ and\ \bibinfo {author} {\bibfnamefont {H.~T.}\ \bibnamefont
  {Quan}},\ }\href {\doibase 10.1103/physreve.105.024101} {\bibfield  {journal}
  {\bibinfo  {journal} {Phys. Rev. E}\ }\textbf {\bibinfo {volume} {105}},\
  \bibinfo {pages} {024101} (\bibinfo {year} {2022})}\BibitemShut {NoStop}%
\end{thebibliography}%

\end{document}